\documentclass[12pt]{article}

\usepackage{feynmp,amsbsy,latexsym}

\newcommand{\no}{\noindent}
\newcommand{\al}{\alpha}
\newcommand{\be}{\beta}

\newcommand{\de}{\delta}
\newcommand{\Ha}{{\cal H}_{\mathrm{Int}}}
\newcommand{\psib}{\bar\psi}
\newcommand{\bd}{{b^{\dag}}}
\newcommand{\dd}{{d^{\dag}}}
\newcommand{\ad}{{a^{\dag}}}
\newcommand{\cd}{\cdot}
\newcommand{\ecd}{{\cdot}}
\newcommand{\eps}{\epsilon}
\newcommand{\mod}[1]{\vert {#1}\vert}
\newcommand{\pa}{\partial}
\newcommand{\G}{{\cal G}}
\newcommand{\tfrac}[2]{{\textstyle\frac{#1}{#2}}}
\newcommand{\ee}{\mathrm{e}}
\newcommand{\phit}{\tilde{\phi}}
\newcommand{\psidirac}{\psi_D}
\newcommand{\psiF}{\psi_f}
\newcommand{\ket}[1]{\vert #1 \rangle}
\newcommand{\ppk}[1]{({#1}+k)^2-m^2+i\eps}
\newcommand{\pmk}[1]{({#1}-k)^2-m^2+i\eps}

\newcommand{\xb}{{\boldsymbol x}}
\newcommand{\yb}{{\boldsymbol y}}
\newcommand{\zb}{{\boldsymbol z}}
\newcommand{\pb}{{\boldsymbol p}}
\newcommand{\qb}{{\boldsymbol q}}
\newcommand{\kb}{{\boldsymbol k}}
\newcommand{\vb}{{\boldsymbol v}}
\newcommand{\intx}{\int\!d^3\xb\;}
\newcommand{\inty}{\int\!d^3\yb\;}
\newcommand{\intp}{\int\!\frac{d^3\pb}{(2\pi)^3}\;}

\newcommand{\intk}{\int\!\frac{d^3\kb}{(2\pi)^3}\;}

\newcommand{\intden}{\int\!\frac{d^4k}{(2\pi)^4}}

\begin{document}

\begin{titlepage}
\rightline{PLY-MS-98-48}
%\leftline{DRAFT (Version printed \today)}
\vskip19truemm
\begin{center}{\Large{\textbf{Charges in Gauge Theories}}}\\ [8truemm]
\textsc{Robin Horan\footnote{email: r.horan@plymouth.ac.uk}, Martin
Lavelle\footnote{email: m.lavelle@plymouth.ac.uk}, and David
McMullan}\footnote{email: d.mcmullan@plymouth.ac.uk}\\ [5truemm]
\textit{School of Mathematics and Statistics\\ The University of
Plymouth\\ Plymouth, PL4 8AA\\ UK} \end{center}

\bigskip\bigskip\bigskip
\begin{quote}
\textbf{Abstract:} In  this  article  we  investigate  charged
particles  in  gauge theories.  After reviewing the physical and
theoretical problems, a method  to  construct  charged  particles
is  presented. Explicit solutions are found in the Abelian theory
and a physical interpretation  is given.  These solutions and our
interpretation of these  variables as  the true  degrees  of
freedom  for  charged particles,  are then tested  in the
perturbative  domain  and are demonstrated   to  yield  infra-red
finite,  on-shell   Green's functions at all orders of
perturbation theory.  The extension to collinear divergences is
studied and it is shown that this method applies  to the case  of
massless  charged  particles. The application of these
constructions to the charged sectors of the standard model  is
reviewed and we conclude  with  a discussion of the successes
achieved so far in this programme and  a list of open questions.
\end{quote}

\bigskip
\begin{quote}
\end{quote}

\medskip

\begin{quote}
 \end{quote} \vfill

\end{titlepage}

\tableofcontents

\setlength{\parskip}{1.5ex plus 0.5ex minus 0.5ex}

\section[Introduction]{Introduction}

%\input{sec1}
%% sec1

One of the most significant advances in particle physics was the
definition of what a particle should be.
Wigner~\cite{Wigner:1939cj} identified a particle with an
irreducible representation of the Poincar\'e group or its covering.
Such representations have a well defined mass and spin. Building
upon free creation and annihilation operators, and
invoking~\cite{weinberg:1995} the cluster decomposition theorem,
one can construct the paradigm quantum field theoretic description
of particles scattering into other particles.

Physical particles generally carry additional quantum numbers such
as isospin, electric and colour charge. These particles are
described by the gauge theories of the standard model. Due to the
masslessness of the gauge bosons, these field theories are plagued
by infra-red divergences which dramatically alter the singularities
of the Green's functions of the matter fields. The states no longer
form irreducible representations of the Poincar\'e
group~\cite{Buchholz:1986uj} and it is therefore widely
believed~\cite{kulish:1970} that there is no particle description
of charges such as the electron. This review will describe the
physics underlying this problem, the manifold consequences of this
breakdown and how a particle description of charged fields may be
recovered.

The standard picture of a particle's journey to a detector is as
follows. When it is shot out of a scattering event it is, after
enough time, a long way away from any other particle and the
interaction between it and the other particles may be neglected.
The free Hamiltonian then describes its dynamics and a particle
description holds. This, the cornerstone of the interaction
picture, breaks down for charged particles interacting via the
gauge theories of the standard model.

The importance of the large distance interactions, which mean
that the residual interactions between particles at large
separations cannot be neglected, is most obvious in Quantum
Chromodynamics (QCD). Experimentalists observe hadrons and
not quarks which are professed to be eternally confined. Although
hadrons are colourless (chargeless as far as the strong nuclear
force is concerned) their masses are rather well described in terms of
building blocks which  are the original Gell-Mann quarks.
However, the current quarks of the QCD Lagrangian are
\emph{not} the constituent quarks of hadronic spectroscopy.
This dichotomy is an unsolved puzzle and shows itself in many
ways. For the light quark flavours the masses of the current
and constituent quarks are very different (roughly two orders of
magnitude for the $u$ and $d$ flavours).
The division of the spin of the proton amongst its constituents
has led to the \lq proton spin crisis\rq, which name shows the
depth of our difficulties in understanding the experimental data.
Finally, since constituent quarks are
presumably constructs made from surrounding a Lagrangian matter
field with a cloud of coloured glue, it is initially at least
highly unclear how the constituents obtain a well-defined colour.

A major aim of the programme of research~\cite{Lavelle:1997ty}
described here is to
understand how constituent quarks arise in QCD, their mutual
interactions and finally how it is that these effective
degrees of freedom are confined.

Electrons are not confined and so it may seem initially less
obvious that the interaction picture paradigm breaks down here
too. Yet the S-matrix elements of Quantum Electrodynamics (QED)
are afflicted by infra-red divergences, as are the on-shell
Green's functions of the theory with external legs corresponding
to charged particles (which we will sometimes generically refer to
as electrons). The infra-red problems change the form of the
Green's functions such that we cannot associate a pole to the
external legs. There are two different sorts of divergence here:
\emph{soft} divergences, which show up in Green's functions and
S-matrix elements, and \emph{phase} divergences which occur in the
phase. These latter divergences are often ignored in QED, but are
of importance in QCD (see Sect.~3.4 of
Ref.~\cite{Ciafaloni:1989vs} and also \cite{DelDuca:1989jt}).

The masslessness of the photon is the underlying cause of the
infra-red problem\footnote{For this reason one sometimes speaks of
mass singularities.}. This vanishing of the mass means that
photons can travel over a large distance and indeed that an
infinite number of soft photons can be created for any finite
amount of energy. Recognition of this led to the Bloch-Nordsieck
answer to the infra-red problem in QED: since any finite
experimental resolution does not restrict the number of photons
which may accompany any charged particle, an experimental
cross-section must come from summing over all these possibilities.
In this sense QED is taken to be a theory defined only at the
level of (measurable) cross-sections and not in terms of
(unobservable) S-matrix elements. Although this is in no way
incompatible with experiment, it is a radical conclusion.

The survival of large distance interactions is responsible for the
claim~\cite{kulish:1970} that charged particles cannot be
incorporated into relativistic quantum field theory. This
conclusion followed from noting that as the usual free Hamiltonian
does not determine the asymptotic dynamics, it must be modified.
This leads to a description in terms of coherent states and there
is no pole structure. Is particle physics then a misnomer?

In fact we will see below that these problems can be resolved
by realising that, as the asymptotic interaction
Hamiltonian is not zero, electric charges are surrounded by an
electromagnetic cloud, just as quarks are by glue. It will then be seen
that such systems
of dressed charges must be gauge invariant. We will
demonstrate that the Green's functions of such dressed charges
are infra-red finite and have a good pole structure.

\bigskip
\subsubsection{General Properties of Charges}
\smallskip

The gauge dependence of the Lagrangian degrees of freedom means
that it is hard to associate any physical meaning to them. This
hinders any attempt to understand how phenomenological models and
concepts can arise from the underlying theory.  As an example of
this we recall that the highly non-trivial gauge dependence of the
Lagrangian vector potential in non-abelian gauge theories has put
difficulties in the way of extracting such, phenomenologically
useful, ideas as effective gluon masses from lattice
calculations~\cite{Henty:1996kv}. Constrained dynamics is the
mathematical tool appropriate to finding gauge invariant degrees of
freedom. The true degrees of freedom then correspond to locally
gauge invariant constructs (which obey Gauss' law). There have been
many attempts to obtain such variables in gauge theories (see,
e.g., \cite{Gogilidze:1997qq} and references therein). We now want
to sketch some general properties that any description of charges
must fulfill.

Local gauge transformations in QED have the form
\begin{equation}\label{gauge}
 A_\mu(x)\to A_\mu(x) + \partial_\mu\theta(x)\,,\qquad
 \mathrm{and}
 \qquad\psi(x)\to \ee^{ie\theta(x)}\psi(x)
  \,,
\end{equation}
so if the coupling $\ee$ could be switched off, then the Lagrangian
fermion would be locally gauge invariant. Similarly Gauss' law
\begin{equation}\label{gausslaw}
  \partial^i F_{i0}=-\ee J_0
\,,
\end{equation}
where $J_0$ is the charge density, would, in the $e=0$ limit, imply
that only the transverse components of the field strength were
physical. However, as we have noted the coupling does not vanish
and so the matter fields cannot be identified with physical
particles. Similarly Gauss' law shows that there is an intimate
link between the matter fields and the electromagnetic cloud which
surrounds them. An immediate consequence of this is that any
description of a charged particle cannot be local since the total
charge can be written as a surface integral at infinity.

We also see that objects which are invariant under global gauge
transformations are chargeless, as the charge density is the
generator of such global transformations. The Gauss law
constraint generates local gauge transformations, so we demand
invariance under local but not global gauge transformations of any
description of a charged particle.

The form of the cloud around a charge determines the
electric and magnetic fields surrounding the charge. This
implies a fundamental non-covariance: the velocity of any charged
particle will determine the nature of the cloud.

The implications of these inevitable properties of
non-locality~\cite{Maison:1975ex} and non-covariance
\cite{Frohlich:1979uu} for \emph{any} description of charged
particles for the general properties of gauge theories have been
investigated both for scattering theory~\cite{Zwanziger:1975ka,
Zwanziger:1975jz} and in axiomatic approaches~\cite{Morchio:1983ym,
Buchholz:1996}.

\bigskip
\subsubsection{Dirac's Dressings}
\smallskip

To the best of our knowledge the interplay between a charged matter field and the
electromagnetic cloud which inevitably surrounds it was first used by
Dirac~\cite{Dirac:1955ca} to construct a specific
description of the electron. He suggested that one should use
\begin{equation}\label{dirac}
 \psidirac(x)\equiv \exp\left( -i\ee\frac{\partial_i A_i}{\nabla^2}\right)\psi(x)
\,.
\end{equation}
This he motivated in the following way: it is locally gauge
invariant\footnote{But not globally gauge invariant.}. Using the
fundamental equal-time commutator,
$[E_i(x),A_j(y)]=i\delta_{ij}\delta(\boldsymbol{x}-
\boldsymbol{y})$, and the representation
\begin{equation}\label{nabladef}
  \frac{\partial_iA_i}{\nabla^2}(x)=
  -\frac1{4\pi}\int\!d^3\boldsymbol{y}
      \frac{\partial_iA_i(x_0,\boldsymbol{y})}{\mod{\boldsymbol{x}-
      \boldsymbol{y}}}
\end{equation}
then the electric field of the state $\psidirac(x)\ket0$ is found
to be
\begin{equation}\label{Efield}
 E^i(x_0,{\boldsymbol{y}})\psidirac(x)\ket0=
-\frac {\ee}{4\pi}
\frac{{\boldsymbol{x}_i-\boldsymbol{y}_i}}{\vert{\boldsymbol{x}-
\boldsymbol{y}}\vert^3}\psidirac(x)\ket0\,,
\end{equation}
which is what one would expect for a static charge.
He further pointed out that this is actually a member of an entire class of
composite fields
\begin{equation}\label{fdress}
   \psiF(x)\equiv \exp\left( -i\ee\int\!d^4zf^\mu(x-z)A_\mu(z) \right)\psi(x)\,,
\end{equation}
which, he argued, are gauge invariant for all $f^\mu$ if it is
demanded that $\pa_\mu f^\mu(w)=\delta^{(4)}(w)$ holds. These
constructs are all evidently non-local. We refer to the cloud
around the matter field as a \emph{dressing}. It is clear that the
dressing suggested by Dirac (\ref{dirac}) is also non-covariant.
This review will be concerned with generalising and refining such
descriptions of charges.

Such pictures have been rediscovered by various authors since Dirac
and there have been many attempts to use certain examples of this
wide class of dressings over the years~\cite{dEmilio:1984,
dEmilio:1984a, Steinmann:1984, Prokhorov:1992, Prokhorov:1993,
Kawai:1995zx,Kawai:1995zv, Kawai:1995zu, Lusanna:1996ut,
Lusanna:1996us, Kashiwa:1997,  Kashiwa:1996, Chechelashvili:1997,
Haagensen:1997pi}.

\medskip

A natural extension of this is to generalise Eq.\ \ref{fdress} to
systems involving more than one matter field. Two opposite
charges at different points can be made gauge invariant by
including a dressing which keeps the entire system gauge invariant.
Such a description could correspond to a positronium state,
a hydrogen atom or a meson.  If the cloud factorises so that
each of the matter fields together with its part of the
dressing is gauge invariant, then we can clearly speak of constituents.
A dressing which does not factorise at all would mean that we had
an effective \lq meson\rq\ field but could not really speak of constituent
particles. The success of the constituent quark picture of hadrons,
taken together with the rough equality of the constituent masses
in mesons and baryons indicates that there must be at least a rough
factorisation of the dressings in some dynamical domains of QCD.

Once one has an ansatz for the dressing, it may be used to study
the interaction between charges. The easiest manner to produce a
gauge invariant description is to link the two matter fields by a
path ordered exponential along some line. The potential is then
obtained\footnote{For details, see Sect.\ 7
of~\cite{Lavelle:1997ty} and~\cite{Haagensen:1997pi}.} by taking
the commutator of the Hamiltonian with this description: that part
of the energy which depends on the separation is the potential.
The string ansatz leads already in QED to a linear, confining
potential. Furthermore the overall coefficient is, as a result of
the infinite thinness of the string, divergent. This string
description has no physical relevance, even in QCD the potential
between two heavy quarks should have a Coulombic form at short
distances~\cite{Cahill:1979dq} and the finite string tension
implies a cigar shaped dressing.

Since the string model corresponds to an (infinitely) excited
state, it is unstable~\cite{Prokhorov:1993}. If we consider two
extremely heavy, fixed charges and neglect pair creation, we may
use the free Hamiltonian and thus solve the time development of the
system starting from the string ansatz initial state. The electric
and magnetic fields immediately broaden out and, of course, lead to
the usual Coulombic field. An animation and detailed discussion of
this can be found at the web site:
\texttt{http://www.ifae.es/\~{}roy/qed.html}

After this fly-by tour of the subject, it is time to get to grips
with the physics of charges. The structure of this article is as
follows. In Sect.\ 2 the form of the interaction Hamiltonian at
large times is investigated. It is shown \emph{not} to vanish for
the matter fields but we demonstrate that, for suitably dressed
fields, the asymptotic interaction does indeed vanish. This result
is then used to construct explicit charged fields. In Sect.~3 an
alternative account of the charged fields in the heavy scalar
theory is presented. These results and our interpretation of them
are then tested in perturbation theory in Sect.\ 4. It is shown
that the soft and phase divergences cancel and a pole structure is
obtained. The massless electron limit is considered in Sect.\ 5 as
a testing ground for the study of collinear divergences. The form
of the asymptotic interaction Hamiltonian is shown to be such that
the solutions of the dressing equation will remove these mass
singularities as well. Subtleties of the dressing equation in this
limit are investigated in this section. Finally Sect.\ 6 reviews
what we have learned and presents a list of outstanding problems.

\section[Charged Particles]{Charged Particles}

As we have seen, the masslessness of the photon implies that the
interactions between charges cannot be \lq switched off\rq\ in the
remote past or future. It is this long range nature of
electromagnetic interactions that lies at the heart of the
infra-red problem in QED. In this section we will review how the
asymptotic dynamics found in such a gauge theory deviates from that
of a free theory. As a consequence of this we will see that the
correct identification of asymptotic, charged particle states can
be made through a process of dressing the matter of the theory.

\subsection[Asymptotic Dynamics]{Asymptotic Dynamics}
The S-matrix codifies the intuitive picture of a scattering
experiment whereby in-coming particles get sufficiently close to
interact, resulting (after the dust settles) in some set of
out-going particles. The in-coming and out-going particle regimes
are identified with elements of the Fock space constructed out of
the creation and annihilation operators found in the free theory.
As discussed in the introduction, the reason for this is that it is
these states that can be identified with (tensor products) of
irreducible representations of the Poincar\'e group. As such, they
are particles!

The key assumption, then, in the S-matrix description of particle
scattering is that in- and out-regimes where the dynamics is that
of the free system exist, i.e., the interacting Hamiltonian must
tend to zero in the remote past and future. It is this assumption
that fails in a gauge theory~\cite{dollard:1964, kulish:1970}.

To see how this arises in QED, we start with the gauge fixed
Lagrangian density:
\begin{equation}\label{qed_lag}
  {\cal L}_{\mathrm{QED}}=-\tfrac14F_{\mu\nu}F^{\mu\nu}+i\psib\gamma^\mu(
  \pa_\mu-i\ee A_\mu)\psi-m\psib\psi+\tfrac12\xi B^2+\pa_\mu A^\mu
  B\,.
\end{equation}
In this we are working in the Lorentz class of gauges (Feynman
gauge corresponding to the choice $\xi=1$). The gauge invariance of the
physical states is encoded in the condition that
$B^{+}|\mathrm{phys}\bigr>=0$. For the present, we do not allow
the mass of the electron to be zero ($m\ne0$).

The interaction Hamiltonian is
given by
\begin{equation}\label{intham}
  \Ha(t)=-e\intx A_\mu(t,\xb)J^\mu(t,\xb)\,,
\end{equation}
where the conserved matter current is
$J^\mu(t,\xb)=\psib(t,\xb)\gamma^\mu\psi(t,\xb)$. In order to
construct the S-matrix, the fields that enter this part of the
Hamiltonian are taken to be in the interaction picture. We recall
that this means that the time evolution of the states is described
by (\ref{intham}) while the fields evolve under the free
Hamiltonian. Thus in (\ref{intham}) we should insert the free field
expansions for the matter and gauge fields, this will then allow us
to study in detail the large $t$ behaviour of the interaction.

We take as our free field expansions
\begin{equation}\label{psi_free}
\psi(x)=\intp\frac1{\sqrt{2E_{\smash{p}}}}\left\{
b(\pb,s)u^s(p)e^{-ip\ecd x}+
\dd(\pb,s)v^s(p)e^{ip\ecd x}
\right\}
\end{equation}
and
\begin{equation}\label{a_free}
A_\mu(x)=\intk\frac1{2\omega_{\smash{k}}}\left\{
a_\mu(\kb)e^{-ik\ecd x}+
\ad_\mu(\kb)e^{ik\ecd x}
\right\} \,,
\end{equation}
where $E_p=\sqrt{|\pb|^2+m^2}$ and $\omega_k=|\kb|$. Inserting
these into (\ref{intham}) results in eight terms which we group
according to the positive and negative frequency components of the
fields. Each of these pieces will have a time dependence of the
form $e^{i\al t}$ where $\al$ involves sums and differences of
energy terms. As $t$ tends to plus or minus infinity, only terms
with $\al$ tending to zero can survive and thus contribute to the
asymptotic Hamiltonian. After performing the spatial integration,
and using the resulting momentum delta function, only  terms of the
form $e^{\pm it(E_{p+k}- E_p\pm\omega_k)}$ have a large $t$-limit;
there are four of them. The requirement that
$E_{p+k}-E_p\pm\omega_k\approx0$ can only be met in QED because the
photon is massless, in which case it implies  that
$\omega_k\approx0$, i.e., only the infra-red regime contributes to
the asymptotic dynamics. From this observation it is
straightforward to see that the full interacting Hamiltonian
(\ref{intham}) has the same asymptotic limit\footnote{A fuller
account of this derivation of the asymptotic interaction will be
presented in Sect.~5 as part of a more general analysis that
includes the case of massless matter.} as
\begin{equation}\label{hint_as}
\Ha^{\mathrm{as}}(t)=-\mathrm{e}\intx
A_\mu(t,\xb)J^\mu_{\mathrm{as}}(t,\xb)
\end{equation}
with
\begin{equation}\label{j_as}
  J^\mu_{\mathrm{as}}(t,\xb)=\intp\frac{p^\mu}{E_p}\rho(p)\delta^3
  \biggl(\xb-\frac{\pb}{E_p} t\biggr)\,.
\end{equation}
The operator content of this current is only contained in the
charge density
\begin{equation}\label{chargedensity}
  \rho(p)=\sum_{s}\Bigl(\bd(p,s)b(p,s)-\dd(p,s)d(p,s)\Bigr)
\end{equation}
which implied that the asymptotic current satisfies the trivial space-time
commutator relation
\begin{equation}\label{jcomm}
[J^\mu_{\mathrm{as}}(x),J^\nu_{\mathrm{as}}(y)]=0\,.
\end{equation}
As such, this current can be interpreted as effectively  the
integral over all momenta of the current associated with a charged
particle moving with velocity $p^\mu/E_p$. Such a current does not
vanish as $t\to\infty$. We thus see that \emph{the asymptotic
dynamics of QED is not that of a free theory.}

The non-triviality of the asymptotic dynamics dramatically alters
the form of the in- and out- matter states. In particular,  their
propagator no longer has a pole-like structure, but instead behaves
near its mass-shell like $(p^2-m^2)^{\be-1}$ where the exponent
$\be$ is gauge dependent; in the Lorentz class of gauges it is
given by~\cite{Bogolyubov:1980nc}
\begin{equation}\label{yennie}
\be=\frac{\ee^2}{8\pi^2}(\xi-3)\,.
\end{equation}
It is this observation that lies at the heart of the
statement~\cite{kulish:1970} that there is no relativistic concept
of a charged particle.

\subsection[Charges as Dressed Matter]{Charges as Dressed Matter}
The persistence  of the asymptotic dynamics in QED means that we
cannot set the electromagnetic coupling to zero for the in-coming
and out-going particle. As an immediate consequence of this we see
that the matter field, $\psi(x)$, cannot be viewed as the field
which creates or annihilates charges since it is not  gauge
invariant in the remote past or future. An equivalent statement of
this fact is that the matter field, $\psi(x)$, does not satisfy
Gauss' law at any time.

To construct a charged field we need to be able to find a
functional of the fields, $h^{-1}(x)$, such that, under a gauge
transformation described by the group element
$U(x)=e^{i\mathrm{e}\theta(x)}$, we have
\begin{equation}\label{h_trans}
  h^{-1}(x)\to h^{-1}(x)U(x)\,.
\end{equation}
Then, from (\ref{gauge}) the charged field will be given by the
gauge invariant product
\begin{equation}\label{cmatter}
 \Psi(x)= h^{-1}(x)\psi(x)\,.
\end{equation}
It is this product which makes precise how we associate charges
with dressed matter: a (chargeless) functional of the fields which
transforms as in  (\ref{h_trans}) is what we mean by a  dressing;
the product (\ref{cmatter}) we then identify as charged matter. We
continue to refer to the matter terms that enter directly into the
Lagrangian  simply as  matter.

There are many distinct ways to construct dressings, reflecting
the fact that our identification (\ref{cmatter}) of charged matter
should be viewed as a minimal requirement. In specific
applications the form of the dressing must be tailored to the
physics at hand: charges that enter into bound states will have a
very different structure to those that describe particle. In our
programme to construct charges explicitly we have so far dealt
exclusively  with the construction of charged
particles\footnote{In non-abelian gauge theories  the gauge field
also needs to be dressed in order for it to carry colour
\cite{Lavelle:1996tz,Lavelle:1994xa,Lavelle:1997ty}.}: it is an
open and immensely interesting problem to extend our approach to
bound states.

In order for the charged matter to describe a particle, we demand
that the dressing is such that  the charged field creates a state
which is an eigenstate of the energy-momentum tensor, i.e., that it
has a sharp momentum. For the matter field, $\psi$, this was not
possible due to the infra-red problem~\cite{Buchholz:1986uj}. For
our charged field, though, we shall see that the form of the
dressing can be chosen  so that there are no infra-red divergences.

 The interacting
Hamiltonian (\ref{intham}) is derived from the coupling term in the
matter part of the Lagrangian density for QED:
\begin{equation}\label{matter_lag}
i\psib(x)\gamma^\mu(\pa_\mu-i\mathrm{e}A_\mu(x))\psi(x)\,.
\end{equation}
We can rewrite
this in terms of the physical charged fields $\Psi(x)$  as
\begin{equation}\label{charged_lag}
i\overline{\Psi}(x)\gamma^\mu(\pa_\mu-i\mathrm{e}A^h_\mu(x))\Psi(x)\,
\end{equation}
where
\begin{equation}\label{ah}
A^h_\mu=h^{-1}A_\mu h+\frac1{i\mathrm{e}}\pa_\mu(h^{-1})h\,,
\end{equation}
which we recognise as a (field dependent) gauge
transformation\footnote{The non-abelian appearance of this gauge
transformation in QED arises from the fact that the dressing is
field dependent and thus might not commute with the potential.} of
the vector potential. Thus, written in terms of the charged fields,
the interacting Hamiltonian is
\begin{equation}\label{chint_ham}
  H^{\mathrm{c}}_{\mathrm{int}}(t)=-\ee\intx A^h_\mu(t,\xb)J^\mu(t,\xb)\,,
\end{equation}
which asymptotically becomes
\begin{eqnarray}\label{chasy_ham}
H^{\mathrm{c}}_{\mathrm{int}}(t)&\to& -\ee\intx
A^h_\mu(t,\xb)J^\mu_{\mathrm{as}}(t,\xb)\\
&=&-\ee\intx\intp\frac{A^h_\mu(t,\xb)p^\mu}{E_p}\rho(p)\delta^3
  \biggl(\xb-\frac{\pb}{E_p} t\biggr)\,.
\end{eqnarray}
This would vanish if we could construct the dressing such that
$A^h_\mu(t,\xb)p^\mu=0$.

To investigate the extent to which this asymptotic interaction can
vanish, we note that the momentum, $p^\mu$, is an arbitrary
on-shell, four vector. Now  through a gauge transformation, such an
algebraic condition can be imposed on the vector potential \emph{at
one point in the mass shell}, but not for the whole mass shell. So
we cannot expect to be able to construct a dressing such that the
asymptotic interaction Hamiltonian (\ref{chasy_ham}) vanishes for
the whole mass-shell. However, we can insist  that our charged
field creates an in-coming or out-going particle with a
\emph{definite momentum}.

From this discussion we see that if we want the charged field
(\ref{cmatter}) to asymptotically create or annihilate a charged
particle moving with four-velocity $u^\mu$, then the dressing must
satisfy the additional kinematical condition that
\begin{equation}\label{dress1}
  u^\mu A^h_\mu(t,\xb)=0\,.
\end{equation}
This will then ensure that, at the point in the mass-shell where
$p^\mu=mu^\mu$, the asymptotic interaction Hamiltonian
(\ref{chasy_ham}) vanishes and thus the state created by the field
will have the appropriate sharp momentum.

What this means in practical terms is that each charged particle
must be constructed out of the matter fields with a different
dressing --- reflecting the  velocity of the particle concerned. At
first sight this looks very peculiar: we are used to particles
being put on-shell as the mass of a particle is a well defined
quantum number, but a particle's velocity is not usually thought of
as a quantum number of the system. For charged particles, though,
velocity is a well defined quantum number. In order not to
interrupt our account of how to construct charges we shall postpone
a discussion of this interesting point until Sect.~3 where the
connection with the heavy charge sector will also be discussed.

\subsection[Construction of Charges]{Construction of Charges}
In order to construct an asymptotic  charged particle moving with
four-velocity $u^\mu$, we need to be able to find a dressing field
$h^{-1}$ such that (\ref{h_trans}) and (\ref{dress1}) hold. The
dressing equation (\ref{dress1}) is, for this purpose, best written
as
\begin{equation}\label{dresseqtn}
  (\eta+v)^\mu\pa_\mu(h^{-1})=-i\ee h^{-1}(\eta+v)^\mu A_\mu\,,
\end{equation}
where we have written the four-velocity $u^\mu$ as
$\gamma(\eta+v)^\mu$, where $\eta^\mu$ is the time like vector
$(1,\boldsymbol{0})$, $v^\mu$ is the space-like vector
$(0,\boldsymbol v)$ with $\boldsymbol v$ the three-velocity of the
charged particle we wish to construct and $\gamma$ is just the
standard relativistic factor $(1-\boldsymbol v^2)^{-\frac12}$.

To see how to construct the dressing, we  first restrict ourselves
to lowest order in the coupling. That is, we take $h^{-1}(x)=1-i\ee
R(x)$ and, to this order, the dressing equation is
\begin{equation}\label{reqtn1}
(\eta+v)^\mu\pa_\mu R(x)=(\eta+v)^\mu A_\mu(x)\,.
\end{equation}
This  equation  can now be easily solved in terms of an integral along the
world line  of a particle moving with the given velocity:
\begin{equation}\label{rsol1}
 R(x)=\int_a^{x^0}(\eta+v)^\mu A_\mu(x_t)dt+\chi(x_a)\,.
\end{equation}
In this  expression we have, for convenience, parametrized the world
line not by the proper time but by $x^\mu_t=x^\mu+(t-x^0)(\eta+v)^\mu$
and we have introduced an arbitrary reference time $a$. The term $\chi(x_a)$ is,
for the moment, an unspecified field configuration in the kernel of
the differential operator $(\eta+v)^\mu\pa_\mu$. It is useful here to
group the $a$-dependent terms together under the integral, and
write $R(x)$ as
\begin{eqnarray}\label{rsol2}
 R(x)&=&\int_a^{x^0}\biggl((\eta+v)^\mu
 A_\mu(x_t)-\frac{d\chi}{dt}(x_t)\biggr)dt+\chi(x)\nonumber\\
 &=& \int_a^{x^0}(\eta+v)^\mu\biggl(
 A_\mu(x_t)-\frac{\pa \chi}{\pa x_t^\mu}(x_t)\biggr)dt+\chi(x)\,.
\end{eqnarray}
In addition to the dressing equation, we also need to ensure the
correct gauge transformation properties for the dressing. To this
order in the coupling, this means that under the gauge
transformation $A_\mu\to A_\mu+\pa_\mu\theta$,
 we must have
\begin{equation}\label{rtrans}
  R(x)\to R(x)+\theta(x)\,.
\end{equation}
This, in turn, implies that
\begin{equation}\label{chitrans}
  \chi(x)\to \chi(x)+\theta(x)\,.
\end{equation}
This is precisely the type  of transformation  that Dirac
investigated in (\ref{fdress}) and we would be tempted to write
\begin{equation}\label{dansatz}
  \chi(x)=\int\!d^4z\,f^\mu(x-z) A_\mu(z)\,,
\end{equation}
where $f^\mu(x-z)$ satisfies $\pa_\mu f^\mu(x-z)=\de^4(x-z)$. But
this only implies (\ref{chitrans}) if \emph{no surface terms arise
when we integrate by parts}. The restriction on the local gauge
transformations to those that vanish at spatial infinity is quite
natural as finite energy restrictions impose a $1/r$ fall-off on
the potential. However, no such restriction applies to the fields
at temporal infinity. Thus the form of $f^\mu(x-z)$ must be such
that no such surface terms arise.

As it stands, we can only infer from this that $f^0(x-z)$ should be
zero outside of some bounded region in the $z^0$-direction. To get
more from this, we make the ansatz that in fact
\begin{equation}\label{ansatz}
  \chi(x)=\int\!d^4z\,G(x-z)\G^\mu A_\mu(z)\,,
\end{equation}
where $\G^\mu$ is a first order differential operator and
$\G\cd\pa\, G(x-z)=\de(x-z)$. In order to avoid the surface terms
that would obstruct the gauge transformation properties of the
dressing, we must have that the operator $\G\cd\pa$
\emph{ cannot involve any time
derivatives.} Given this restriction, we see that
$G(x-z)=\de(x^0-z^0)G(\xb-\zb)$. We shall, for convenience, write
$\chi$ as
\begin{equation}\label{chinotation}
  \chi(x)=\frac{\G\cd A}{\G\cd\pa}(x)\,.
\end{equation}

The temporal parameter $a$ which enters into the form of the
dressing (\ref{rsol1}) has been introduced by hand and thus should
not directly affect any physical results. Taking the derivative of
(\ref{rsol2}) with respect to $a$ yields the gauge invariant term
\begin{equation}\label{aderiv1}
  -(\eta+v)^\mu(A_\mu(x_a)-\pa_\mu \chi(x_a))\equiv
  -(\eta+v)^\mu\biggl(A_\mu(x_a)-\pa_\mu \frac{\G^\nu
  A_\nu}{\G{\cd}\pa}(x_a)\biggr)\,.
\end{equation}
This we can write in a form where the gauge invariance is manifest
as
\begin{equation}\label{aderiv2}
  -(\eta+v)^\mu\biggl(\frac{\G^\nu
  \pa_\nu A_\mu}{\G{\cd}\pa}(x_a)- \frac{\G^\nu\pa_\mu
  A_\nu}{\G{\cd}\pa}(x_a)\biggr)
  =-(\eta+v)^\mu\frac{\G^\nu F_{\nu\mu}}{\G{\cd}\pa}(x_a)\,.
\end{equation}
The  condition selecting physical states is that they are
annihilated by the $B$-field, thus any physical observable must
commute with $B(x)$. The relevant equation of motion that follows
from (\ref{qed_lag}) is, to this order, $\pa^\nu
F_{\nu\mu}=\pa_\mu B$. This tells us that, in order for
(\ref{aderiv2})  to act trivially on physical states, it must be
equal to
\begin{equation}\label{aderiv3}
  (\eta+v)^\mu\frac{\pa^\nu F_{\nu\mu}}{\G{\cd}\pa}(x_a)\,.
\end{equation}
This now allows us to find the form for $\G^\mu$ and hence the
dressing to this order.

The first order operator $\G^\mu$ is constructed out of the
vectors $\pa^\mu$, $n^\mu$ and $v^\mu$ that characterise the
theory. The anti-symmetry of the field strength $F_{\nu\mu}$ in
(\ref{aderiv2}) implies that (\ref{aderiv3}) will follow if
\begin{equation}\label{g1}
  \G^\nu=-\pa^\nu+(\eta+v)^\nu\bigl(\al(n{\ecd}\pa)+\be(v{\ecd}\pa)\bigr)\,.
\end{equation}
In which case the second order operator $\G\cd\pa$ is
\begin{equation}\label{gpa}
  \G\cd\pa
        =\nabla^2+\be(v\ecd\pa)^2+(\al-1)(\eta\ecd\pa)^2+(\al+\be)(v\ecd\pa)
        (\eta\ecd\pa)\,,
\end{equation}
where $\nabla=\pa_i\pa_i$. We have seen that it is essential for
the gauge invariance of the charges that this operator has no
time derivatives in it. This follows only if $\al=1$ and $\be=-1$.
Hence we see that
\begin{equation}\label{g2}
  \G^\mu=(\eta+v)^\mu(n-v)\ecd\pa-\pa^\mu\,.
\end{equation}
Thus, to lowest order in the coupling, we have seen that $h^{-1}(x)=1-
i\mathrm{e}R(x)$
where $R(x)$ is the sum of two terms:
\begin{equation}\label{horder1}
  R(x)=-\int_a^{x^0}\!ds\,(\eta+v)^\nu\frac{\pa^\mu F_{\mu\nu}}{\G\cd\pa}(x_t)
  +\frac{\G^\mu A_\mu}{\G\cd\pa}(x)\,.
\end{equation}
As will become apparent when we perform perturbative test of our
construction, this decomposition of the dressing into two terms
reflects the two manifestations of the infra-red in QED that were
discussed in the introduction: the soft divergence and the phase
divergence. Indeed, the first (gauge invariant) part of $R(x)$ will
be shown to be responsible for controlling the  phase structure of
charged matter; while the second term will be responsible for the
soft dynamics of the charged particle.

It is, perhaps, helpful to specialise to the static situation in
order to get a better feel for the form of the dressing we have
been constructing. In that case $v=0$ and we get
\begin{equation}\label{rstatic}
  R(x)=-\int_a^{x^0}\!ds\,\frac{\pa^i F_{i0}}{\nabla^2}(s,\xb)
  +\frac{\pa_i A_i}{\nabla^2}(x)\,,
\end{equation}
where $1/\nabla^2$ has been defined in (\ref{nabladef}). The
expected spatial non-locality of the charge is now manifest. More
generally, the inverse to $\G\cd\pa$ is given by
\begin{equation}\label{nonlocalv}
  \frac1{\G\cd\pa}f(t,\xb)=-\frac{1}{4\pi}\inty\frac{f(t,\yb)}{\|\xb-\yb\|_v}\,,
\end{equation}
where
\begin{equation} \label{vdef}
\frac1{\|\zb\|_v}:=
\frac1{2\pi^2}\int\! d^3\kb\,\frac{e^{i \kb\ecd\zb}}
{V^v\cd k}\,,
\end{equation}
and $V^v_\mu=(\eta+v)_\mu(\eta-v)\cd k-k_\mu$.

In order to go beyond this first order result we must take into
account the fact that the fields which enter into the dressing are
operators, and thus we need to know their commutation properties.
Equal time commutators between the fields can be directly read off
from the Lagrangian (\ref{qed_lag}). However, the non-locality
displayed in (\ref{horder1}) means that we need general space-time
commutators between the fields. For the fully interacting theory
there is no general approach to finding these commutators without
first solving the whole theory. Charges as particles, though,  only
make physical sense in the asymptotic regime governed by the
interacting Hamiltonian (\ref{hint_as}). In this regime, the
relevant  dynamics of the gauge field is described by the simpler
set of equations:
\begin{eqnarray}\label{eqofmotion}
\pa^\nu F_{\nu\mu}&=& \pa_\mu B-\ee J_\mu^{\mathrm{as}}\\
\pa^\mu A_\mu&=&-\xi B\,,
\end{eqnarray}
where the matter coupling is now through the asymptotic current
(\ref{j_as}). For this theory the space-time commutators can be
explicitly constructed, as we will now show.

In Feynman gauge ($\xi=1$) the potential satisfies
\begin{equation}\label{boxaj}
  \Box A_\mu=-\ee J^{\mathrm{as}}_\mu\,.
\end{equation}
The general solution to this is
\begin{equation}\label{sola}
   A_\mu(x)=A_\mu^{\mathrm{free}}(x)-\ee\int\!d^4yD_{R}(x-y)J^{\mathrm{as}}_\mu(y)\,,
\end{equation}
where $D_{R}(x-y)$ is the retarded Green's function for the $\Box$
operator and $A_\mu^{\mathrm{free}}$ is a solution of the
homogeneous equation $\Box A_\mu^{\mathrm{free}}=0$. From the form
of this solution for the potential, and given the triviality of the
asymptotic current commutations (\ref{jcomm}), the space-time
commutator for the potential in this interacting theory are seen to
be the same as that for the free theory.

The free field equation $\Box A_\mu^{\mathrm{free}}=0$  implies
that
\begin{equation}\label{atrick}
 A_\mu^{\mathrm{free}}(x)=\int\!d^3z\bigl(\pa_0^zD(x-z)
 A_\mu^{\mathrm{free}}(z)-D(x-z)\pa_0A_\mu^{\mathrm{free}}(z)\bigr)
\end{equation}
where $D(x-z)$ is the commutator function for free fields:
\begin{equation}\label{commfn}
  D(x-y)
  =-\intk\frac1{\omega_k}e^{i\kb\ecd(\xb-\yb)}\sin
  \Bigl(\omega_k(x^0-y^0)\Bigr)\,.
\end{equation}
The identification in (\ref{atrick}) is made by first observing
that the right hand side is independent of $z^0$: setting
$z^0=x^0$ then implies the result. Exploiting this
$z^0$-independence, the commutator
$[A_\mu^{\mathrm{free}}(x),A_\nu^{\mathrm{free}}(y)]$ is simply
calculated by using (\ref{atrick}) with $z^0=y^0$. Then the equal
time commutation relations
$[A_\mu(y),\dot{A}_\nu(z)]_{\mathrm{et}}=-ig_{\mu\nu}\de(\yb-\zb)$
can be used. This, in conjunction with our observation that the
free and asymptotic interacting fields have the same commutators,
results   in the space time commutators in Feynman gauge being:
\begin{equation}\label{acomm}
  [A_\mu(x),A_\nu(y)]=[A_\mu^{\mathrm{free}}(x),A_\nu^{\mathrm{free}}(y)]=-ig_{\mu\nu}D(x-y)\,.
\end{equation}
Using this, the first order result (\ref{horder1}) can be
exponentiated (while still preserving the $a$-independence of the
construction on physical states) to give
\begin{equation}\label{hexpo}
  h^{-1}(x)=e^{-i\ee K(x)}e^{-i\ee \chi(x)}\,,
\end{equation}
where
\begin{equation}\label{softbitofh}
\chi(x)=\frac{\G^\mu A_\mu}{\G\cd\pa}(x)\,,
\end{equation}
and
\begin{eqnarray}\label{phasebitofh}
K(x)&=&-\int_a^{x^0}ds\,(\eta+v)^\mu\frac{\pa^\nu
F_{\nu\mu}}{\G\cd\pa}(x_s)+ \ee\int_{-\infty}^a ds (\eta+v)_\mu
\frac{J^{\mathrm{as}\mu}}{\G\cd\pa}(x_s)\\
&&\qquad\qquad\qquad-\tfrac12\ee\gamma^{-1}u\ecd
x\intk\frac1{\kb^2-(\kb\ecd\vb)^2}
\end{eqnarray}
These results are, by construction,  independent of $a$. Setting
$a=-\infty$ gives (modulo the field independent, tadpole term)
\begin{equation}\label{phaseinh}
  K(x)=-\int_{-\infty}^{x^0}ds\,(\eta+v)^\mu\frac{\pa^\nu
F_{\nu\mu}}{\G\cd\pa}(x_s)\,,
\end{equation}
which is a form adapted to performing perturbative calculations.
The full derivation of these results will be presented elsewhere.

In summary, we have seen that charged fields corresponding to
charged particles moving with momentum, $p^\mu=m u^\mu$ may be
described by
\begin{equation}\label{hexponentiated}
  \Psi_p(x)=e^{-i\ee K_p(x)}e^{-i\ee \chi_p(x)}\psi(x)\,,
\end{equation}
where the dressing satisfies both the gauge transformation
property~(\ref{h_trans}) and the dressing
equation~(\ref{dresseqtn}) with the four velocity appropriate for
the particular momentum~$p$. Our construction has shown that the
on-shell Green's functions of these dressed fields, taken at the
correct points on their mass shells, should not suffer from any
infra-red divergences. Such formal arguments must be tested: in
Sect.~4 we will show that this prediction~\cite{Lavelle:1997ty} is
true to all orders in perturbation theory~\cite{Bagan:1998kg}.

We now conclude this construction of charges in QED with an
observation and a comment. Note that in the static limit the charge
field is given by
\begin{equation}\label{diracdressing}
  \exp\left(i\ee\int_{-\infty}^{x^0}\!ds
  \frac{\pa^iF_{i0}}{\nabla^2}(s,\xb)\right)
  \exp\left(-i\ee\frac{\pa_i A_i}{\nabla^2}\right)\psi(x)=
  \exp\left(i\ee\int_{-\infty}^{x^0}\!ds
  \frac{\pa^iF_{i0}}{\nabla^2}(s,\xb)\right)\psidirac(x)
\end{equation}
where $\psidirac$ was Dirac's proposal (\ref{dirac}) for a static
charged field based on the form of the electric field produced by
such a charge. The additional term in (\ref{diracdressing}), that
follows from the more fundamental dressing equation approach
presented here, does not alter the form of the electromagnetic
field and was thus missed by Dirac. Its role in the cancellation of
the phase divergence will be presented in Sect.~4.

The construction of the charge above has been done in Feynman
gauge. It is essential to verify that the results hold for the full
Lorentz class of gauges. While it is true that the very
construction of the charges (\ref{cmatter}) ensures gauge
invariance, it is not immediately clear that this is equivalent to
the independence of the construction from the gauge fixing
parameter $\xi$. In particular, changing $\xi$ from one will modify
the form of the space-time commutators, (\ref{acomm}), and thus
could have a potentially non-trivial impact on the construction of
the dressing. A full proof of the gauge invariance of these dressed
fields will be presented elsewhere, however, in Sect~4, we shall
demonstrate perturbatively that the $n$-point Green's functions
constructed out of the charged fields are manifestly
$\xi$-independent. Before making these tests, we will now present
an alternative approach to the physical restrictions  on dressings.

\section[Heavy Charges]{Heavy Charges}

We have seen in the last section that the construction of the
dressing for a charged particle depends on the velocity of the
charge. In this section we wish to discuss, in more general terms,
the role  the velocity of a  charged particle plays  in the
structure of the states of QED . We will then look at heavy
charges, where velocity is also singled out, and see how   the
asymptotic regime,  governed by the asymptotic Hamiltonian
(\ref{hint_as}), can be characterised in terms of the mass of the
charge. In addition, we will also see a further derivation for the
dressing equation (\ref{dresseqtn}).

\subsection[Moving Charges]{Moving Charges}
It is not possible \cite{Haag:1992hx} to talk about the space of
states of QED without first identifying the observables of the
theory and their algebra. The dual requirements of gauge invariance
and locality on any physical observable have immediate and quite
striking implications for the structure of QED.

A familiar example of this is the fact that charge is
superselected: the state space is a direct sum of different charge
sectors and the action of any physical observable cannot change the
charge. In order to derive this physically important result we
recall that Gauss' law (\ref{gausslaw}) implies that on physical
(gauge invariant) states there is a precise relation between the
electric field and the charge density. In particular, the total
charge $Q$ is given on such states by the spatial integral
\begin{equation}\label{chargeq}
  Q=\intx \pa_iE_i(t,\xb)\,.
\end{equation}
Using Gauss' theorem we can relate this expression for the charge
to a surface integral and, in particular, to one at spatial
infinity. That is, if we define the electric flux in the direction
$\hat{\xb}$ by
\begin{equation}\label{flux}
\mathcal{E}_i(
\hat{\xb}):=\lim_{R\to\infty}
R^2E_i(\xb+R\hat{\xb})\,,
\end{equation} then the total
charge is given by the \lq
flux at infinity\rq\ through
\begin{equation}\label{qinfinity}
Q=\int_{S^2_\infty}
d\boldsymbol{s}\cdot\boldsymbol{\mathcal{E}}\,.
\end{equation} The
superselection properties of the charge then follow immediately
from the obvious fact that any local observable will always commute
with the flux at spatial infinity.

This simple argument can be extended to show why velocity plays
such a central role in the identification of charged particles.
Recall that the the electric field of a charge whose present
position is $y$ and  moving with velocity $\vb$ is
\begin{equation}\label{movinge}
 E_i(\xb)=-\frac{\ee\gamma}{4\pi}
\frac{(x-y)_i}{\left(\gamma^2\Bigl((\vb\cdot(\xb-\yb)\Bigr)^2
+|\xb-\yb|^2\right)^{3/2}}\,.
\end{equation}
The electric flux at spatial infinity is then
\begin{equation}\label{fluxv}
\mathcal{E}_i(\hat{\xb})=-\frac{\ee\gamma}{4\pi}
\frac{\hat{x}_i}{\left(\gamma^2(\vb\cdot\hat{\xb})^2
+1\right)^{3/2}}
\end{equation}
Just as above for the total charge, this flux will commute with all
local observables and thus can be used to characterise distinct
sectors of the theory. As this flux only depends on the velocity of
the charge we see that, in the asymptotic regime where charged
particles exist, their velocity is a well defined quantum number.

Within the class of local observables, this means that velocity is
superselected. However, the very non-locality of the construction
of the charges implies that we have to extend the algebra of
observables to include  non-local constructions such as dressings.
As we will see in the next section, the perturbative construction
of Green's functions for charged matter will involve non-local
interactions between different charges and thus allow  charges with
different velocities to interact. This does not mean, though, that
the total charge $Q$ is no-longer superselected: the dressings are
all chargeless and hence the interactions they induce commute with
the charge.

\subsection[Heavy Matter]{Heavy Matter}
In our discussion of the asymptotic interactions found in QED, we
have simply taken the large $t$-limit of the interaction
Hamiltonian in the interaction picture. A disadvantage of studying
this limit, though, is that it does not supply a description of the
scale for the onset of the asymptotic dynamics and thus of the
domain within which velocity is a valid quantum number. More
properly, then, we should investigate the limit as some
dimensionless parameter gets large. In QED with massive matter, the
natural parameter is the product of $t$ and the mass scale of the
theory set by $m$, the lightest mass of the system. In this sense,
the dynamics at large time is equivalent to that of a theory with
heavy charges. Given the central role of the asymptotic dynamics in
our programme, it is important to understand precisely how the
asymptotic interactions emerge in the heavy sector, and to
understand the significance of the dressing equation for heavy
matter.

It is well known~\cite{weinberg:1995} that the infra-red structures
found in QED are independent of the spin of the charged particles,
thus it is instructive to see how the interaction (\ref{hint_as})
also emerges from the heavy sector of scalar~QED. The matter part
of the QED Lagrangian is now
\begin{equation}\label{scalarmatter}
  (D_\mu\phi)^\dagger(D^\mu\phi)-m^2\phi^\dagger\phi\,.
\end{equation}
The heavy limit~\cite{Georgi:1990um} can only be taken at specific
points on the mass-shell of the particle. To this end, one
introduces the rescaled fields
\begin{equation}\label{phitilde}
 \phit(x):=\sqrt{2m}e^{imu\ecd x}\phi(x)\,,
\end{equation}
where we have chosen a four-velocity $u^\mu$ ($u\cd u=1$) that will
ultimately describe the velocity of the heavy charge. In terms of
these new fields, the matter part of the Lagrangian becomes
\begin{equation}\label{scalarmatter2}
  i\phit^\dagger u^\mu D_\mu\phit+\frac1{2m}(D_\mu\phit)^\dagger(D^\mu\phit)\,.
\end{equation}
In the large $m$-limit only the first term survives and the
equations of motion for the heavy matter become
\begin{equation}\label{hmeqm}
  u^\mu D_\mu\phit=0\,.
\end{equation}
The interaction Hamiltonian is easily identified in this limit and
is constructed out of the current
\begin{equation}\label{heavyj}
  J^\mu_{\mathrm{heavy}}(x)=u^\mu\phit^\dagger(x)\phit(x)\,,
\end{equation}
which has, as expected, precisely the same form as the asymptotic
current (\ref{j_as}) for the specific point on the mass-shell
described by $u^\mu$.

Heavy charged matter also needs to be constructed out of the heavy
matter through the process of dressing. Thus the heavy scalar
charge is given by the gauge invariant field
\begin{equation}\label{hcharge}
  \tilde\Phi(x)=h^{-1}(x)\phit(x)\,,
\end{equation}
where the dressing is that appropriate to a charge moving with the
given four velocity $u^\mu$:
\begin{equation}\label{heavyde}
  u^\mu\pa_\mu(h^{-1})=-i\ee h^{-1} u^\mu A_\mu\,.
\end{equation}
This dressing equation, in conjunction with the equation of motion
for the heavy matter (\ref{hmeqm}) implies that
\begin{equation}\label{hceqm}
  u^\mu \pa_\mu\tilde\Phi=0\,.
\end{equation}
That is, the state created by the field $\varphi=e^{-imu\ecd
x}\tilde\Phi$ is an eigenstate of the momentum operator:
\begin{equation}\label{eigenstate}
  P_\mu\varphi|0\rangle=mu_\mu\varphi|0\rangle\,,
\end{equation}
i.e., as claimed, it is a particle with momentum $mu^\mu$.

Having seen that the heavy sector characterises the regime
described by the asymptotic Hamiltonian (\ref{hint_as}), we can now
identify the onset of the asymptotic dynamics, and thus the
emergence of a charged particle interpretation, with the
kinematical regime where the lightest mass $m$ of the matter
dominates the momentum flow in any process. This is precisely the
domain of the eikonal approximation in soft photon
amplitudes~\cite{sterman:1993} whereby the photon has a
spin-independent coupling to the matter, which acts as an external
point-like source. It is also the domain where scattering theory
applies and we will now test our construction.

\section[Perturbation Theory]{Perturbation Theory}

%\input{sec4}

%% The section

\noindent Our
aim now is to show how we may apply the physical,  dressed fields
we derived  in the last section  to remove infra-red  divergences
already \textit{at  the level of Green's functions}. We will start
with a one-loop example, where we will classify the various  sorts
of  mass  singularities  present  in  an unbroken Abelian  gauge
theory.   Then we will demonstrate~\cite{Bagan:1998kg} their
explicit cancellation  when physical, dressed fields are used.
Finally we will show that this cancellation is in fact general and
occurs at all orders.  In this review we will solely concern
ourselves with the   IR-structure   and   not   with   the
renormalisation   of UV-singularities  ---  full  calculations  of
the renormalisation constants associated with the physical
propagators  of spinor and scalar   QED  were,  however,   given in
Ref.'s~\cite{Bagan:1997su} and~\cite{Bagan:1997dh} respectively.
Ref.'s~\cite{weinberg:1995, muta:1987, sterman:1993} offer
introductions to infra-red divergences in perturbation theory.

\subsection[One-Loop Calculations]{One-Loop Calculations}
%\smallskip
The full spectrum  of infra-red  divergences  that  characterise
gauge theories where the charged particles  are massive all occur
in pair creation. We thus start by briefly reviewing this process
for undressed matter fields. We will use the theory of scalar
QED since these divergences are insensitive to the spin of the
particles and we may so avoid the technical complications of Diracology.
The Feynman rules are

%\smallskip
\begin{fmffile}{fig41}
\[
    \parbox{20mm}{\begin{fmfgraph*}(20,10)
      \fmfleft{in} \fmfright{out}
      \fmf{fermion,lab=$p$}{in,out}
      \end{fmfgraph*}}\;
      = {\displaystyle \frac{i}{p^2-m^2}} \,,\quad
      \parbox{18mm}{\begin{fmfgraph*}(18,10)
      \fmfleft{in} \fmfright{out}
      \fmf{photon,lab=$k$}{in,out}
      \fmfv{lab=$\mu$,lab.angle=90}{in}
      \fmfv{lab=$\nu$,lab.angle=90}{out}
      \end{fmfgraph*}} \;
      ={\displaystyle \frac{-ig_{\mu\nu}}{k^2}}\,,\quad
      \parbox{18mm}{\begin{fmfgraph*}(18,20)
      \fmfleft{upp} \fmfright{in,out}
      \fmf{fermion,lab=p}{in,vtx}
      \fmf{fermion,lab=q}{vtx,out}
      \fmf{photon}{vtx,upp}
      \fmfv{lab=$\mu$,lab.angle=90}{upp}
      \end{fmfgraph*}}\;
      =i\ee(p+q)^\mu
\]
\end{fmffile}
\noindent\textbf{Fig.\       4.1:}      \textsl{The       Feynman
rules.}

\medskip
We have chosen to work in Feynman gauge, but the physical results will
be shown to be gauge invariant below. The  seagull  ($\phi^{*}\phi  AA$)
vertex does not   alter  the spin  independent  IR-behaviour  and
is therefore irrelevant to our purpose.

\bigskip
\subsubsection[One-Loop Pair Creation]{One-Loop Pair Creation}
\smallskip
Consider  now pair creation from a classical  source\footnote{The
Feynman  rule for the vertex with the source  is here taken to be
unity for simplicity,  it could also be given the renormalisation
group invariant form, $m^2\phi^{*}\phi$.} as shown in Fig.~4.2.

\begin{fmffile}{fig42}
\[
    {{\begin{fmfgraph*}(30,25)
      \fmfleft{source} \fmfright{pone,ptwo}
      \fmf{dashes,tension=1.5}{source,ver}
      \fmf{vanilla}{ver,v1}
   \fmfv{label=$q$,label.angle=-80,label.dist=.1w}{v1}
      \fmf{vanilla}{ver,v2}
   \fmfv{label=$p$,label.angle=80,label.dist=.1w}{v2}
      \fmf{fermion}{v1,pone}
      \fmf{fermion}{v2,ptwo}
      \fmffreeze
      \fmf{photon,tension=1}{v1,v2}
    \end{fmfgraph*}}}
    \]
\end{fmffile}
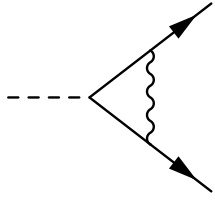
\noindent\textbf{Fig.\ 4.2:} \textsl{Covariant pair production
diagram.}

\medskip

The vertex  function  corresponding  to Fig.\ 4.2  is given  in
Feynman gauge by
\begin{equation}\label{gamone}
\Gamma=
\frac{-i\ee^2}{[p^2-m^2][{q}^2-m^2]}
\intden \frac{(2q+k)^\mu(2p-k)^\nu}{[(q+k)^2-m^2+i\eps][
(p-k)^2-m^2+i\eps]}\frac{g_{\mu\nu}}{k^2+i\eps}
\,.
\end{equation}
Naive  power  counting  tells  us that  this  does  not have  an
infra-red  divergence  when the outgoing particles are off-shell.
However,  if we extract  a simple  pole for each  of our outgoing
particles,   then  the  on-shell   residue  may  be  seen  to  be
IR-divergent
\begin{equation}\label{resone}
\Gamma_{\hbox{\scriptsize IR}}=
\frac{i\ee^2}{[p^2-m^2][{q}^2-m^2]}
\intden \frac{{q}^\mu p^\nu}{(q\cdot k+i\eps)
(p\cdot k-i\eps)}\frac{g_{\mu\nu}}{k^2+i\eps}
\,,
\end{equation}
where we have dropped higher powers of $k$ as they do not lead to
IR-divergences  (this  means  that we are henceforth  restricting
ourselves to loop momenta smaller than some cutoff).  The easiest
way   to  calculate   these   divergences   is  to  perform   the
$k_0$~integral using Cauchy's theorem. The four poles are:
\begin{equation}
k_0=\pm\mod{\boldsymbol{k}} \mp i\eps\,,\qquad
k_0={\boldsymbol{q}}\cdot {\boldsymbol{k}}/q_0 -i\eps\,,\qquad
k_0={\boldsymbol{p}}\cdot {\boldsymbol{k}}/p_0 +i\eps\,.
\end{equation}
The results  of the two different  sets  of poles  have different
physical interpretations.   To see this distinction,  let us take
them  one at a time.

The contribution from the $k^2$ poles to (\ref{resone}) is easily
found to be
\begin{equation}\label{restwo}
\frac{\ee^2}{(p^2-m^2)({q}^2-m^2)}\frac1{8\pi^2}\frac1{\mod{
\boldsymbol{v}_{\hbox{\scriptsize rel}}}}\log\left(\frac{1+\mod{
\boldsymbol{v}_{\hbox{\scriptsize rel}}}}
{1-\mod{\boldsymbol{v}_{\hbox{\scriptsize rel}}}}\right)
\int \frac{d\mod{\boldsymbol{k}}}{\mod{
\boldsymbol{k}}}
\,,
\end{equation}
where $\boldsymbol{v}_{\hbox{\scriptsize rel}}$  is the relative
velocity  between the two
charged particles~\cite{weinberg:1995}.  In (\ref{restwo}) there is
a logarithmic  divergence  in  the  small  $\mod{\boldsymbol{k}}$
limit, i.e., there is a divergent contribution  from soft virtual
photons. This is called a \textit{soft} divergence.

The  other  poles  are  often  neglected  as they  correspond  to
divergences in the (unobservable)  phase of the Green's function,
which  cancel  in  cross-sections.   These  structures  yield  in
(\ref{resone})
\begin{equation}
\frac{-\ee^2}{(p^2-m^2)({q}^2-m^2)}    \frac1{4\pi^2}   \int^1_{-1}
\frac{du}{\mod{\boldsymbol{v}_{\hbox{\scriptsize rel}}}u-i\eps}
    \int
\frac{d\mod{\boldsymbol{k}}}{\mod{\boldsymbol{k}}}
\,.
\end{equation}
An IR-divergence is again visible. To perform the $u$-integral we
can now employ the relation
\begin{equation}\label{dlt}
\frac1{u-i\eps}=\mbox{\textsf{PV}}\frac1u +i\pi \delta(u)\,,
\end{equation}
where \textsf{PV} denotes the principle value. Only the last term
in (\ref{dlt}) contributes  as the other leads to an odd integral
in $u$. We so obtain
\begin{equation}\label{resthree}
\frac{-i\ee^2}{(p^2-m^2)({q}^2-m^2)}\frac1{4\pi}\frac1{\mod{
\boldsymbol{v}_{\hbox{\scriptsize rel}}}} \int
\frac{d\mod{\boldsymbol{k}}}{\mod{\boldsymbol{k}}}
\,.
\end{equation}
The extra factor of $i$ here, compared to (\ref{restwo}), betrays
the fact that this divergence occurs in the (unobservable) phase.
Such singularities are called \textit{phase} divergences. (Note
that  this  is why we chose  a pair
creation process:  if we had taken a scattering process then both
of the poles in ${\boldsymbol{p}}_i \cdot {\boldsymbol{k}}$ would
be in the same half-plane and could be neglected.) Now we
want to demonstrate  that the use of the correct, gauge invariant
fields removes both of these divergences.

\bigskip
\subsubsection[Perturbation       Theory       with       Dressed
Fields]{Perturbation Theory with Dressed Fields}
\smallskip

 Using dressed fields
means calculating  Green's  functions  of the fields  given in
(\ref{hexponentiated}).  Since  the dressings explicitly  depend on
the coupling,  $\ee$, we must also expand the dressings  as well as
including  the usual interaction  vertices. Thus  we introduce  new
vertices  and hence  new  diagrams.   The diagrammatic  rules for
dressed  Green's functions  are then just the standard ones
augmented by the extra vertices which come from expanding  the
dressings  in powers  of the  coupling.   The two factors in the
dressing  each yield a different  vertex structure.  The Feynman
rules corresponding to the dressings are given in Fig.~4.3.

\smallskip
\begin{fmffile}{fig43}
\[
 \parbox{20mm}{\begin{fmfgraph*}(20,15)
   \fmfleft{i}\fmfright{o}\fmftop{g}\fmf{fermion,label=$p- k$}{i,o}
   \fmf{photon,tension=0,label=$\searrow^{{\displaystyle
   k}}$,label.side=left}{g,o}
   \fmflabel{$\mu$}{g}
   \fmfv{decor.shape=circle,decor.filled=empty,decor.size=2mm}{o}
 \end{fmfgraph*}}
\quad ={\displaystyle \frac{\ee V_p^\mu}{V_p\cdot k}}\,,\qquad\quad
 \parbox{20mm}{\begin{fmfgraph*}(20,20)
   \fmfleft{i}\fmfright{o}\fmftop{g}\fmf{fermion,label=$p- k$}{i,o}
\fmf{photon,tension=0,label=$
      \searrow^{{\displaystyle k}}$,label.side=left}{g,o}
   \fmflabel{$\mu$}{g}
   \fmfv{decor.shape=cross,decor.size=3mm}{o}
 \end{fmfgraph*}}\quad
 ={\displaystyle \frac{\ee W^\mu_p}{V_p\cdot k}} \,.
\]
\end{fmffile}

\noindent\textbf{Fig.\       4.3:}      \textsl{The       Feynman
rules from expanding the dressings. The first vertex comes from
the soft ($\chi$) part of the dressing, and the latter corresponds to the
phase ($K$) term.}

\medskip
\no The dressings are, of course, dependent upon the momentum of the particle
being dressed and so here we have defined
\begin{equation}\label{vwdef}
  V^\mu_p:= (\eta+v)^\mu(\eta-v)\cd k - k^\mu
  \,,\quad
  W_p^\mu:= \frac{(\eta+v)\cd k k^\mu - (\eta+v)^\mu k^2}{k\cd(\eta+v)}
\end{equation}
where $v=(0,\boldsymbol{v})$ is the velocity of a particle with
momentum $p=m\gamma(1, \boldsymbol{v})$.

Since the dressed  fields are, by construction,  gauge invariant,
so are their (connected) Green's  functions.   In the connected
vertex  function,  there  are  two further  diagrams  with  extra
interaction  vertices, Fig.'s 4.4.b-c.  The remaining diagrams of
Fig.~4.4 come from expanding both parts of the dressing.

\begin{fmffile}{fig44}
\[
\begin{array}{ccc}
    {{\begin{fmfgraph*}(30,25)
      \fmfleft{source} \fmfright{pone,ptwo}
      \fmf{dashes,tension=1.5}{source,ver}
      \fmf{vanilla}{ver,v1}
   \fmfv{label=$
   q$,label.angle=-80,label.dist=.1w}{v1}
      \fmf{vanilla}{ver,v2}
   \fmfv{label=$
   p$,label.angle=80,label.dist=.1w}{v2}
      \fmf{fermion}{v1,pone}
      \fmf{fermion}{v2,ptwo}
      \fmffreeze
      \fmf{photon,tension=1}{v1,v2}
    \end{fmfgraph*}}\atop{(a)}} &
       {{\begin{fmfgraph}(30,25)
      \fmfleft{source} \fmfright{pone,ptwo}
      \fmf{dashes}{source,ver}
      \fmf{vanilla}{ver,u1,u2}
      \fmf{vanilla}{u2,pone}
      \fmf{vanilla}{ver,v1,v2}
      \fmf{vanilla}{v2,ptwo}
      \fmffreeze
      \fmf{photon,left}{v1,v2}
    \end{fmfgraph}}\atop{(b)}} &
 {{\begin{fmfgraph}(30,25)
      \fmfleft{source} \fmfright{pone,ptwo}
      \fmf{dashes}{source,ver}
      \fmf{vanilla}{ver,u1,u2}
      \fmf{vanilla}{u2,ptwo}
      \fmf{vanilla}{ver,v1,v2}
      \fmf{vanilla}{v2,pone}
      \fmffreeze
      \fmf{photon,left}{v1,v2}
    \end{fmfgraph}}\atop{(c)}}     \\ [5mm]
     {{\begin{fmfgraph}(30,25)
         \fmfleft{source} \fmfright{pone,ptwo}
      \fmf{dashes}{source,ver}
      \fmf{vanilla}{ver,u1,u2}
      \fmf{vanilla}{u2,u3}
       \fmf{vanilla,tension=3}{u3,ptwo}
      \fmf{vanilla}{ver,v1,v2}
      \fmf{vanilla}{v2,v3}
      \fmf{vanilla,tension=3}{v3,pone}
      \fmffreeze
      \fmf{photon,right=0.15}{pone,ptwo}
      \fmfv{decor.shape=circle,decor.filled=empty,decor.size=2mm}{pone}
      \fmfv{decor.shape=circle,decor.filled=empty,decor.size=2mm}{ptwo}
    \end{fmfgraph}}\atop{(d)}}
     &
   {{\begin{fmfgraph}(30,25)
         \fmfleft{source} \fmfright{pone,ptwo}
      \fmf{dashes}{source,ver}
      \fmf{vanilla}{ver,u1,u2}
      \fmf{vanilla}{u2,u3}
       \fmf{vanilla,tension=3}{u3,ptwo}
      \fmf{vanilla}{ver,v1,v2}
      \fmf{vanilla}{v2,v3}
      \fmf{vanilla,tension=3}{v3,pone}
      \fmffreeze
      \fmf{photon,right}{u2,ptwo}
      \fmfv{decor.shape=circle,decor.filled=empty,decor.size=2mm}{ptwo}
    \end{fmfgraph}}\atop{{(e)}}}
      &
 {{\begin{fmfgraph}(30,25)
         \fmfleft{source} \fmfright{pone,ptwo}
      \fmf{dashes}{source,ver}
      \fmf{vanilla}{ver,u1,u2}
      \fmf{vanilla}{u2,u3}
       \fmf{vanilla,tension=3}{u3,ptwo}
      \fmf{vanilla}{ver,v1,v2}
      \fmf{vanilla}{v2,v3}
      \fmf{vanilla,tension=3}{v3,pone}
      \fmffreeze
      \fmf{photon,left}{v2,pone}
      \fmfv{decor.shape=circle,decor.filled=empty,decor.size=2mm}{pone}
    \end{fmfgraph}}\atop{(f)}}
       \\ [5mm]
        {{\begin{fmfgraph}(30,25)
         \fmfleft{source} \fmfright{pone,ptwo}
      \fmf{dashes}{source,ver}
      \fmf{vanilla}{ver,u1,u2}
      \fmf{vanilla}{u2,u3}
       \fmf{vanilla,tension=3}{u3,ptwo}
      \fmf{vanilla}{ver,v1,v2}
      \fmf{vanilla}{v2,v3}
      \fmf{vanilla,tension=3}{v3,pone}
      \fmffreeze
      \fmf{photon,right=0.15}{pone,ptwo}
      \fmfv{decor.shape=cross,decor.size=3mm}{pone}
      \fmfv{decor.shape=cross,decor.size=3mm}{ptwo}
    \end{fmfgraph}}\atop{(g)}}
     &
   {{\begin{fmfgraph}(30,25)
         \fmfleft{source} \fmfright{pone,ptwo}
      \fmf{dashes}{source,ver}
      \fmf{vanilla}{ver,u1,u2}
      \fmf{vanilla}{u2,u3}
       \fmf{vanilla,tension=3}{u3,ptwo}
      \fmf{vanilla}{ver,v1,v2}
      \fmf{vanilla}{v2,v3}
      \fmf{vanilla,tension=3}{v3,pone}
      \fmffreeze
      \fmf{photon,right}{u2,ptwo}
      \fmfv{decor.shape=cross,decor.size=3mm}{ptwo}
    \end{fmfgraph}}\atop{{(h)}}}
      &
 {{\begin{fmfgraph}(30,25)
         \fmfleft{source} \fmfright{pone,ptwo}
      \fmf{dashes}{source,ver}
      \fmf{vanilla}{ver,u1,u2}
      \fmf{vanilla}{u2,u3}
       \fmf{vanilla,tension=3}{u3,ptwo}
      \fmf{vanilla}{ver,v1,v2}
      \fmf{vanilla}{v2,v3}
      \fmf{vanilla,tension=3}{v3,pone}
      \fmffreeze
      \fmf{photon,left}{v2,pone}
      \fmfv{decor.shape=cross,decor.size=3mm}{pone}
    \end{fmfgraph}}\atop{(i)}}
       \\ [5mm]
  {{\begin{fmfgraph}(30,25)
         \fmfleft{source} \fmfright{pone,ptwo}
      \fmf{dashes}{source,ver}
      \fmf{vanilla}{ver,u1,u2}
      \fmf{vanilla}{u2,u3}
       \fmf{vanilla,tension=3}{u3,ptwo}
      \fmf{vanilla}{ver,v1,v2}
      \fmf{vanilla}{v2,v3}
      \fmf{vanilla,tension=3}{v3,pone}
      \fmffreeze
      \fmf{photon,right=0.15}{pone,ptwo}
  \fmfv{decor.shape=circle,decor.filled=empty,decor.size=2mm}{pone}
      \fmfv{decor.shape=cross,decor.size=3mm}{ptwo}
    \end{fmfgraph}}\atop{(j)}}
     &
     {{\begin{fmfgraph}(30,25)
         \fmfleft{source} \fmfright{pone,ptwo}
      \fmf{dashes}{source,ver}
      \fmf{vanilla}{ver,u1,u2}
      \fmf{vanilla}{u2,u3}
       \fmf{vanilla,tension=3}{u3,ptwo}
      \fmf{vanilla}{ver,v1,v2}
      \fmf{vanilla}{v2,v3}
      \fmf{vanilla,tension=3}{v3,pone}
      \fmffreeze
      \fmf{photon,right=0.15}{pone,ptwo}
      \fmfv{decor.shape=cross,decor.size=3mm}{pone}
 \fmfv{decor.shape=circle,decor.filled=empty,decor.size=2mm}{ptwo}
    \end{fmfgraph}}\atop{(k)}}
      &
       \\
\end{array}
\]
\end{fmffile}

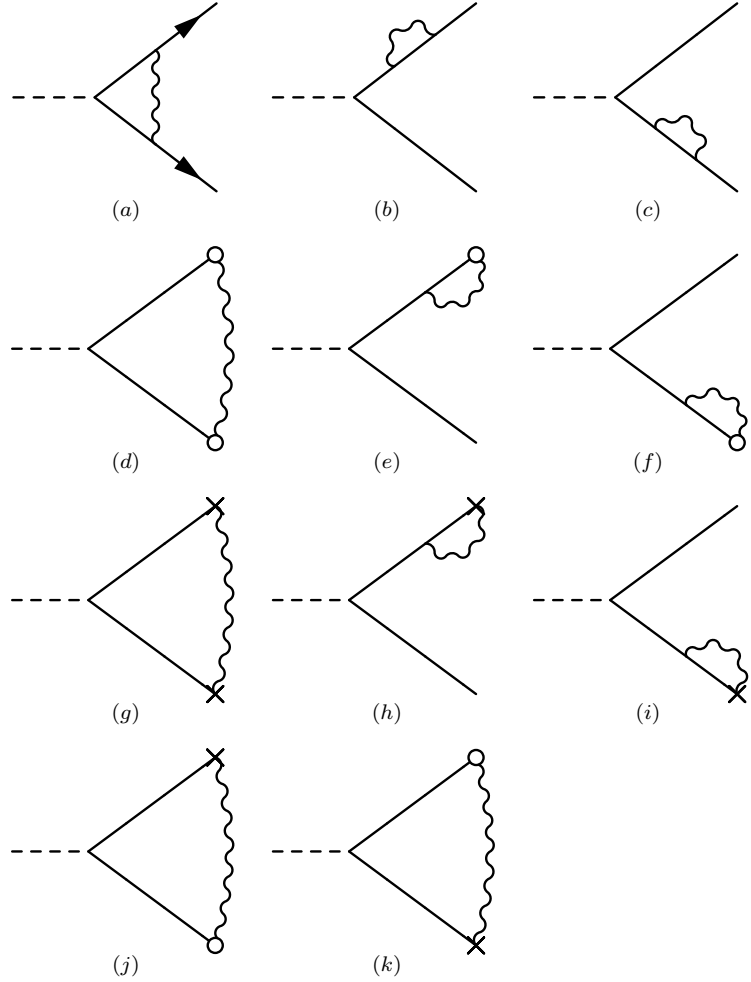
\noindent\textbf{Fig. 4.4:} \textsl{All one-loop Feynman diagrams
in  the  pair  creation  process  which  contain  IR-divergences.
Diagrams   $(a)-(c)$   are  covariant;   $(d)-(f)$   involve  the
perturbative  expansion  of  the  $\chi$~term  in  the  dressing;
$(g)-(i)$ comes from expanding the $K$~term; finally the diagrams
$(j)$  and $(k)$  are cross  terms  from expanding  both dressing
structures.}

\medskip

Our procedure  is to extract  the IR-divergences  in the on-shell
residue after we have taken out a pole for each external leg. For
simplicity   we  work  in  Feynman  gauge,  however,   the  gauge
invariance  of  our  dressed  Green's  functions  will  be  fully
apparent  in  our  final  results.  We  will  employ  dimensional
regularisation  since  it  preserves  gauge  invariance.  We  may
neglect the sorts of diagram shown in Fig.~4.5: those of type (a)
are   massless   tadpoles   (and   so   vanish   in   dimensional
regularisation),  type  (b) do not yield  IR-divergences  when we
extract  the poles, which reflects  the spin independence  of the
infra-red structure,  and diagrams of type (c) are also found not
to yield IR-divergences in the residue.

\bigskip
\begin{fmffile}{fig45}
\begin{displaymath}
    {{\begin{fmfgraph}(30,25)
      \fmfleft{source} \fmfright{pone,ptwo}
      \fmf{dashes,tension=1.5}{source,ver}
      \fmf{fermion}{ver,pone}
      \fmf{fermion}{ver,v3}
\fmf{phantom, tension=20}{v3,ptwo}
      \fmffreeze
\fmf{photon,tension=0.9,right=90}{v3,v3}
      \fmfv{decor.shape=square,decor.filled=empty,decor.size=2mm}{v3}
      %\fmfforce{(0.8w,h)}{v3}
    \end{fmfgraph}}\atop{(a)}} \qquad
       {{\begin{fmfgraph}(30,25)
      \fmfleft{source} \fmfright{pone,ptwo}
      \fmf{dashes}{source,ver}
      \fmf{vanilla}{ver,v2}
      \fmf{fermion}{v2,pone}
      \fmf{vanilla}{ver,v1}
      \fmf{fermion}{v1,ptwo}
      \fmffreeze
      \fmf{photon,right=90,tension=0.9}{v1,v1}
    \end{fmfgraph}}\atop{(b)}} \qquad
    {{\begin{fmfgraph}(30,25)
      \fmfleft{source} \fmfright{ptwo,pone}
      \fmf{dashes}{source,ver}
      \fmf{vanilla}{ver,v2}
      \fmf{fermion}{v2,pone}
      \fmf{vanilla}{ver,v1}
      \fmf{fermion}{v1,ptwo}
      \fmffreeze
      \fmf{photon}{pone,v1}
       \fmfv{decor.shape=square,decor.filled=empty,decor.size=2mm}{pone}
    \end{fmfgraph}}\atop{(c)}}
\end{displaymath}
\end{fmffile}

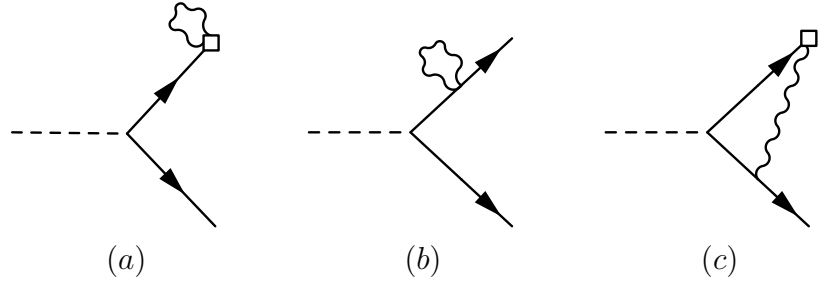
\noindent\textbf{Fig.\ 4.5:} \textsl{Classes of diagrams
which do not contribute IR-divergences to the residue. The square
vertex here and below signifies that the generic contributions of
both parts of the dressing are meant.}

\medskip

Let us now consider the contribution  from the $\chi$ part of the
dressing.   Since the other term in the dressing  is itself gauge
invariant, this part, taken together with the covariant diagrams,
must be gauge  invariant.   The diagrams  4.4.e-f  have two poles
already,  but Fig.~4.4.d  appears  not to have any poles:  it is,
however, already IR-divergent even off-shell. This has nothing to
do with the usual IR-divergences  which, we recall, arise when we
go on-shell. We can, however, extract poles from such diagrams as
we now describe.

\bigskip
\subsubsection[Factorisation]{Factorisation}
\smallskip
This  sort of off-shell  IR-divergence  appears  when one or more
photons are exchanged  between two dressings.   We will call such
diagrams  \textsl{rainbow}   diagrams.   When  we  extract  poles
associated  with the external  legs,  we find the residue  has an
\textit{on-shell} IR-divergence. Together with the other diagrams
of Fig.~4.4, the divergences  from the rainbow diagrams will then
(if our predictions of the IR-finiteness  of the physical Green's
functions  is correct)  cancel the soft and phase divergences  of
the covariant  (dressing independent)  diagrams.   This procedure
makes much use of the following algebraic identity:
\begin{equation}\label{ident}
\frac1{(p-k)^2-m^2}=\frac1{p^2-m^2}\left[1+
\frac{2p\cdot k-k^2}{(p-k)^2-m^2}\right]
\,.
\end{equation}
The  integrand  from  the  Feynman  rules  for Fig.~4.4.d  may be
thus rewritten as (we do not write superfluous factors)
\begin{eqnarray}\label{rnone}
& & \!\!\!\!\!\!  \!\!\!\!\!\! \intden
\frac{V_p\cdot V_{q}}{V_p\cdot k\,V_q\cdot k\, k^2}
\frac1{[\ppk p][\pmk{q}]} \nonumber \\ &
=&  \!\!\!\!
\intden \frac{V_p\cdot V_{q}}{V_p\cdot k\,V_q\cdot k\, k^2}
\frac1{p^2-m^2}\frac1{\pmk q}\left[1-\frac{2p\cdot
k+k^2}{\ppk p}
\right]\!.
\end{eqnarray}
The first  term in the square  bracket  here is the only relevant
one. Before we consider it, let us show that the second term does
not  contribute   to  the  residue:    extracting   the  pole  in
$1/(q^2-m^2)$ we obtain from this term
\begin{equation}
\frac{1}{[p^2-m^2][q^2-m^2]}
\intden \frac{V_p\cdot V_{q}}{V_p\cdot k\,V_q\cdot k \,k^2}
\frac{k^2-2p\cdot k}{\ppk p}
\left[1+\frac{2q\cdot k-k^2}{\pmk q}
\right]
\,.
\end{equation}
We have extracted the two poles and may now go on-shell:  but the
square  bracket  here,  and hence the contribution  of this second
term,  vanishes  on-shell.    Returning now  to  (\ref{rnone})   the
contribution of the first term is easily found to be
\begin{equation}\label{fall}
\frac1{[p^2-m^2][q^2-m^2]}
\intden \frac{V_p\cdot V_{q}}{V_p\cdot k\,V_q\cdot k\, k^2}
\left[1+\frac{2q\cdot k-k^2}{\pmk q}
\right]
\,.
\end{equation}
However,  the one in the square  bracket  here is just a massless
tadpole and thus vanishes. On-shell the second term becomes $-1$
and we obtain  the final  result  that,  as far as IR-divergent
terms in the residue are concerned, we have
\begin{eqnarray}
\intden &&\!\!\!\!\!\! \!\!\!\!\!\!
\frac{V_p\cdot V_{q}}{V_p\cdot k\,V_q\cdot k\, k^2}
\frac1{[\ppk p][\pmk{q}]} \nonumber \\
&
\to& \frac1{[p^2-m^2][q^2-m^2]}\times (-1)
\intden \frac{V_p\cdot V_{q}}{V_p\cdot k\,V_q\cdot k\, k^2}
\,,
\end{eqnarray}
i.e., the rainbow line has been stripped off the original
integral to yield a factor of
\begin{equation}\label{Cdefn}
C_{pq}=-\intden \frac{V_p\cdot V_{q}}{V_p\cdot k\,V_q\cdot
k\, k^2}
\,,
\end{equation}
times the Feynman rules for the diagram without this line.  It may be
demonstrated  that   this   property   of  the
\textit{factorisation}  of rainbow  lines is completely  general.
For  a diagram  with  $n$  rainbow  lines,  as far  as  the  soft
divergences  in the on-shell  residue of the poles are concerned,
\emph{the  rainbow   lines  may  be  stripped   off  and  replaced
by}
$(C_{pq})^n$.  A diagrammatic proof of the factorisation property
will be published elsewhere.   We will make extensive use of this
property in the all orders proof below.

\bigskip
\subsubsection[Tadpoles]{Tadpoles}
\smallskip
With the factorisation  property we are able to calculate  all of
the one-loop  diagrams.   However,  it seems worthwhile  to first
remark on the identification  of IR- and UV-singularities  since,
as is well known, in dimensional  regularisation  we can exchange
IR- and UV-divergences  because massless tadpoles  vanish in this
scheme.  We have used this cancellation  above in (\ref{fall}).
Using this identity the badly defined, soft, off-shell divergences
were removed  and replaced  by terms which  yield IR~divergences
only when we went on shell.  These on-shell singularities,  as we
will soon show, help to  cancel the soft on-shell  divergences  of
the other  diagrams.   This  replacement  of  ill-defined,
off-shell singularities  in  rainbow  diagrams  is the  only  place
in our calculations  where  we drop tadpoles.   At no stage  do we
alter structures which first develop soft or phase divergences only
on-shell.

However,  there are other massless  tadpole  diagrams  associated
with dressed Green's functions, as well as the tadpoles  which we
dropped  above  in Eq.~\ref{fall}.   Let  us now  consider  their
interplay --- firstly for the dressed propagator.  Fig.~4.6 shows
the generic diagrams which enter here:

\bigskip
\begin{fmffile}{fig46}
\begin{displaymath}
    {{\begin{fmfgraph}(30,25)
      \fmfleft{in} \fmfright{out}
      \fmf{vanilla}{in,v1}
      \fmf{vanilla}{v1,v2}
      \fmf{vanilla}{v2,out}
      \fmffreeze
      \fmf{photon,left}{v1,v2}
 \end{fmfgraph}}\atop{(a)}} \qquad
       {{\begin{fmfgraph}(30,25)
       \fmfleft{in} \fmfright{out}
      \fmf{vanilla}{in,v1}
      \fmf{vanilla}{v1,out}
      \fmffreeze
       \fmfv{decoration.shape=square,decoration.filled=empty,decor.size=2mm}{in}
      \fmf{photon,left=.75}{in,v1}
 \end{fmfgraph}}\atop{(b)}} \qquad
  {{\begin{fmfgraph}(30,25)
       \fmfleft{in} \fmfright{out}
      \fmf{vanilla}{in,out}
        \fmffreeze
       \fmfv{decoration.shape=square,decoration.filled=empty,decor.size=2mm}{in}
     \fmfv{decoration.shape=square,decoration.filled=empty,decor.size=2mm}{out}
      \fmf{photon,left=.5}{in,out}
      \end{fmfgraph}}\atop{(c)}} \qquad
      {{\begin{fmfgraph}(30,25)
       \fmfleft{in} \fmfright{out}
      \fmf{phantom,tension=20}{in,v1}
      \fmf{vanilla}{v1,out}
          \fmffreeze
         \fmf{photon,tension=1,right=90}{v1,v1}
       \fmfv{decoration.shape=square,decoration.filled=empty,decor.size=2mm}{v1}
      \end{fmfgraph}}\atop{(d)}}
\end{displaymath}
\end{fmffile}

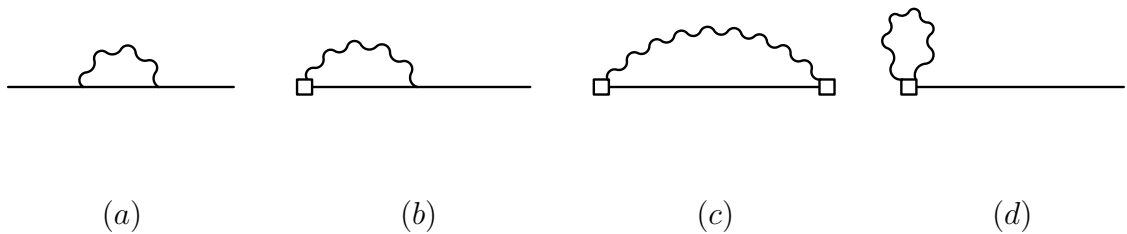
\noindent\textbf{Fig.\ 4.6:} \textsl{The classes of diagrams in
the one-loop dressed propagator.}

\medskip

\no Here  there  are  massless
tadpole  diagrams, Fig.\ 4.6.d,  and  rainbow  diagrams,  Fig.~4.6.c.  The
propagator has, of course, no phase divergence, so let us solely consider the
contribution of the soft part of the dressing here\footnote{It is easy to
check that the phase part of the dressing does not
bring in any soft divergences}. Using
(\ref{ident})  to  strip  off  the  rainbow  line  we obtain from this
diagram (for the soft part of the dressing) the massless tadpole integral

\begin{equation}\label{klapsch}
\frac{i\ee^2}{p^2-m^2}\intden\frac{V_p^2}{(V_p\cdot
k)^2}\frac{-i}{k^2}
\,,
\end{equation}
which we can drop in dimensional  regularisation.   However,  the
contribution  of the two massless tadpole diagrams, of the type
shown in Fig.~4.6.d, is  rapidly  found  from  the  Feynman   rules
to  be  equal  to (\ref{klapsch})  but with  the opposite  sign and
so cancels  it exactly\footnote{Note  that  the  massless tadpole
diagrams  of Fig.~4.5.b  do not cancel,  but they, of course, do
not alter the on-shell dependence of the IR-structure}. The
remaining integrals conspire  to yield an infra-red  finite
propagator  as has been shown elsewhere in
detail~\cite{Bagan:1997su, Bagan:1997dh}.

This exact cancellation  does not carry through to higher Green's
functions.  In general a gauge invariant subset of tadpoles survives.
This appears to be a subtlety concerned with taking products of
distributions. It has been suggested that such problems may perhaps be
removed by using the Hertz formulation of QED~\cite{dEmilio:1984}.
The removal of these tadpoles in a general fashion is a worthy
object for further study, here we content ourselves with
noting that they vanish in dimensional regularisation and that we do not
drop tadpoles connected with on-shell infra-red divergences.

\bigskip
\subsubsection[Cancellation of the Soft and Phase
Divergences]{Cancellation of the Soft and Phase Divergences}
\smallskip
Putting together  the one-loop  diagrams relevant to the soft
structure, i.e., \hbox{Fig.'s~4.4.a-f}, extracting a pole for each
of the external  legs and going on-shell,  we obtain   the
following terms  in the residue which are by power counting
IR-divergent:
\begin{eqnarray}\label{indone}
I_{\hbox{\scriptsize IR}}=
\intden
\Bigg\{\Bigg. && \!\!\!\!\!\! \!\!\!\!
\left(\frac{{p}^\mu}{p\cdot k}-\frac{{V_q}^\mu}{{V_q}\cdot k}\right)
\frac{g_{\mu\nu}}{k^2}
\left(\frac{{V_p}^\nu}{V_p\cdot k}-\frac{{q}^\nu}{{q}\cdot k}\right)
\nonumber \\
&& \quad -
\left(\frac{{p}^\mu}{p\cdot k}-\frac{{q}^\mu}{{q}\cdot k}\right)
\frac{g_{\mu\nu}}{k^2}
\left(\frac{{p}^\nu}{p\cdot k}-\frac{{q}^\nu}{q\cdot k}\right)
\Bigg.\Bigg\}
\,,
\end{eqnarray}
where  we have removed  a factor  of $i\ee^2/[(p^2-m^2)(q^2-m^2)]$.
This is a gauge invariant  result:   replacing  the Feynman gauge
propagator by \textit{any}  more general form necessarily  brings
in  a  factor   of  $k_\mu$   or  $k_\nu$   which   vanishes   on
multiplication into the brackets in (\ref{indone}). Similar gauge
invariant  structures  are obtained  in the study  of the dressed
propagator and in other physical vertex functions. We now have to
show that the soft divergences seemingly apparent in (\ref{indone})
in fact cancel.

\bigskip
The cancellation  may be demonstrated  by performing  the integrals
explicitly,  using the methods  outlined  above.  However,  it is
simpler to realise that
\begin{equation}
\frac{V_p^\mu}{V_p\cdot k}=
\frac{(\eta+v)^\mu(\eta-v)\cdot k -k^\mu}{{\boldsymbol{k}}^2
-({\boldsymbol{k}}\cdot{\boldsymbol{v}})^2}
\,,
\end{equation}
which, at the soft pole, $k^2=0$, we may write as
\begin{equation}
\frac{V_p^\mu}{V_p\cdot k}=
\frac{(\eta+v)^\mu(\eta-v)\cdot k -k^\mu}{(k\cdot \eta)^2
-({\boldsymbol{k}}\cdot{\boldsymbol{v}})^2}
\,.
\end{equation}
We may drop  the $k^\mu$  term  in the numerator  here  (cf,  the
argument  for  the  gauge  invariance  of  Eq.~\ref{indone}).  So
effectively we have
\begin{equation}
\frac{V_p^\mu}{V_p\cdot k}=
\frac{(\eta+v)^\mu(\eta-v)\cdot k}{(k\cdot \eta)^2
-({\boldsymbol{k}}\cdot{\boldsymbol{v}})^2}
=\frac{(\eta+v)^\mu}{(\eta+v)\cdot k}
=\frac{p^\mu}{p\cdot k}
\,,
\end{equation}
where it is important to note that we have taken $p$ to be at the
correct  point  on the mass  shell,  $p=m\gamma(\eta+  v)$, which
defines the appropriate dressing. (If we were to go on-shell at a
different  point,  this  equality  would  not  hold  and the soft
divergences do not cancel.)

In all cases of gauge invariant  Green's  functions,  we are thus
able to replace
\begin{equation}
\frac{V_p^\mu}{V_p\cdot k}=
\frac{p^\mu}{p\cdot k}\,,\qquad \hbox{\rm and}\qquad
\frac{V_q^\mu}{V_q\cdot k}=
\frac{q^\mu}{q\cdot k}
\,,
\end{equation}
as far  as the  soft  divergences  are  concerned.   \textsl{This
equality  makes  the  cancellation  of the  soft  divergences  in
(\ref{indone})  and other dressed  Green's functions  immediately
apparent.}

\bigskip

Up till now we have not included  the other gauge invariant  part
of  the  dressing.  It  is  straightforward  to  check  that  the
contribution  of this structure  to the pair creation  process is
gauge  invariant.   It may also be demonstrated  that it does not
spoil the above cancellation  of soft divergences.   (Some of the
individual  diagrams  which  involve  this  part of the dressing,
Fig.'s~4.4.g-k,   contain  soft  divergences,  but  overall  they
cancel.) The role of this structure  is, we expect, to cancel the
phase  divergence  of  Fig.~4.4.a.    The  only  phase  divergent
contribution from the entire dressing comes from a single diagram
involving only the $K$-structure:   Fig.~4.4.g.  From the Feynman
rules we obtain from this figure after factorisation

\begin{equation}
\intden \frac{W_p^\mu}{V_p\cd k} \frac{W_q^\nu}{V_q\cd k}
\frac{g_{\mu\nu}}{k^2} = \intden
\frac{(\eta+v)\cd(\eta+v')k^2}{V_p\cd k V_q\cd k [(\eta+v)\cd
k-i\epsilon][(\eta+v')\cd k +i\epsilon]} \,,
\end{equation}
where     we    have     not     written     the    two     poles
$-i\ee^2/([p^2-m^2][q^2-m^2])$  and have thrown away terms which do
not yield phase divergences.   This result  may be readily  seen to
cancel exactly with (\ref{resthree})  when we go on-shell  at the
right point.

\medskip

To summarise the one-loop results:  the dressed Green's functions
are, by construction,  gauge invariant.  If we go on-shell at the
points  on the mass shells such that the momenta  of the external
lines and the velocities that define the various dressings agree,
then  the soft  and phase  divergences  cancel.   We now want  to
show that this holds at all orders.

\subsection[All Orders]{All Orders}
\smallskip
\noindent    The   physical   dressed   propagator    was   shown
elsewhere~\cite{Bagan:1997su,  Bagan:1997dh}  to be IR-finite  at one loop.
However,  we can show that  this holds  at all orders  through  a
slight     extension     of    the    work    of    Jackiw    and
Soloviev~\cite{soloviev:1968}.

These authors used low energy theorems and spectral representations
to study the behaviour of the propagator around the pole,
$p^2=m^2$, in various gauges. In covariant gauges they regained the
usual result~(\ref{yennie}), while in non-covariant gauges there is
a pole if the following integral vanishes (this is their Eq.\ 3.41)
\begin{equation}\label{jacsol}
F_R=\frac{\ee^2}{(2\pi)^3}\int^4k e^{ik\ecd k}\theta(k_0) \delta(k^2)
\frac{r_\mu r_\nu}{(r\ecd k)^2} \Pi^{\mu\nu}_R\,,
\end{equation}
here $r$ corresponds to the mometum of the matter field and $\Pi$
is the photon propagator up to a factor of $i/k^2$. Now our
dressed propagator, for a charge with momentum $r=m\gamma(\eta,\boldsymbol{v})$
corresponds to the usual matter propagator in
the gauge where
\begin{equation}
 \Pi^{\mu\nu}_R =
 -g^{\mu\nu} +\frac{k^\mu (\eta+v)^\nu + (\eta+v)^\mu k^\nu }{k\ecd(\eta+v)}
 -\frac{k^\mu k^\nu}{(k\ecd(\eta+v))^2} \left[ V_r^2+2V_r\cd
 k\right]\,.
\end{equation}
taking the $\delta(k^2)$ factor of the integrand into account, it
is easy to see that at the right point on the mass shell (\ref{jacsol})
vanishes and the propagator has a pole. This was pointed out
explicitly for the static case (Coulomb gauge) in
Ref.~\cite{soloviev:1968}.

\bigskip
\subsubsection[The Dressed Propagator]{The Dressed Propagator}
\smallskip

Since  we  now  know  that  the  dressed propagator has a pole, we see that
the wave  function   renormalisation
constants  are  IR-finite. We can therefore multiply  the dressed  Green's
functions by them without introducing new IR-divergences. We will
use  this  freedom  to show  the cancellation  of soft  and phase
singularities  in dressed Green's functions  at all orders.   Our
approach  is diagrammatic.   First we shall consider the types of
diagrams which can introduce IR-divergences in the propagator, in
this  way  we can describe  the structure  of the  wave  function
renormalisation constants.

The non-dressed  propagator,  i.e., just the covariant  diagrams,
can be represented at all orders in the following way:

\begin{fmffile}{ufig41}
\[ %\begin{center}
  \parbox{40mm}{\begin{fmfgraph}(40,15)
      \fmfleft{in}\fmfright{out}
      \fmf{vanilla,tension=2}{in,j}
      \fmf{vanilla}{j,jj}
      \fmf{vanilla,tension=3}{jj,v1}
      \fmf{vanilla}{v1,u}
       \fmf{vanilla,tension=2}{u,v2}
      \fmf{dashes}{v2,v3}
      \fmf{vanilla,tension=2}{v3,v4}
      \fmf{vanilla}{v4,out}
   \fmfv{decor.shape=circle,decor.filled=shaded,decor.size=4mm}{v1,u,v4}
    \end{fmfgraph}} \quad
=:\frac i{p^2-m^2-\Sigma}\,.
\] %\end{center}
\end{fmffile}

\noindent
When  we take  dressings  into  account,  these  chains  must  be
supplemented   by  various  possible  structures   which  we  now
enumerate. The first is

\begin{fmffile}{ufig42}
%%%%%%%%%%%%%%%%% Defining brown_muck style %%%%%%%
\fmfcmd{%
vardef port (expr t, p) =
 (direction t of p rotated 90)
  / abs (direction t of p)
 enddef;}

\fmfcmd{%
  vardef portpath (expr a, b, p) =
   save l; numeric l; l = length p;
    for t=0 step 0.1 until l+0.05:
    if t>0: .. fi point t of p
      shifted ((a+b*sind(180t/l))*port(t,p))
     endfor
    if cycle p: .. cycle fi
 enddef;}

\fmfcmd{%
  style_def brown_muck expr p =
   shadedraw(portpath(thick/2,2thick,p)
    ..reverse(portpath(-thick/2,-2thick,p))
    ..cycle)
  enddef;}
%%%%%%%%%%%%%%%%% end of style %%%%%%%%%%%%%%%%
\[
  2\,\parbox{40mm}{\begin{fmfgraph}(40,15)
      \fmfleft{in}\fmfright{out}
      \fmf{vanilla,tension=2}{in,j}
      \fmf{vanilla}{j,jj}
      \fmf{vanilla,tension=3}{jj,v1}
      \fmf{vanilla}{v1,u}
       \fmf{vanilla,tension=2}{u,v2}
      \fmf{dashes}{v2,v3}
      \fmf{vanilla,tension=2}{v3,v4}
      \fmf{vanilla}{v4,v5}
      \fmf{brown_muck}{v5,out}
      \fmfv{decor.shape=circle,decor.filled=shaded,decor.size=4mm}{v1,u,v4}
    \end{fmfgraph}}
\quad =:\frac{2i\tilde\Sigma}{p^2-m^2-\Sigma}\,.
\]
\end{fmffile}

\smallskip
\noindent where the 2 accounts for the possibility  that dressing
corrections, $\tilde\Sigma$, can be indistinguishably attached at
either end in the scalar theory.

There  may of course  also be such dressing  corrections  at both
ends. This is diagrammatically
\begin{fmffile}{ufig43}
  %%%%%%%%%%%%%%%%% Defining brown_muck style %%%%%%%
\fmfcmd{%
vardef port (expr t, p) =
 (direction t of p rotated 90)
  / abs (direction t of p)
 enddef;}

\fmfcmd{%
  vardef portpath (expr a, b, p) =
   save l; numeric l; l = length p;
    for t=0 step 0.1 until l+0.05:
    if t>0: .. fi point t of p
shifted  ((a+b*sind(180t/l))*port(t,p))  endfor  if cycle  p:  ..
cycle fi enddef;}

\fmfcmd{%
  style_def brown_muck expr p =
   shadedraw(portpath(thick/2,2thick,p)
    ..reverse(portpath(-thick/2,-2thick,p))
    ..cycle)
  enddef;}
%%%%%%%%%%%%%%%%% end of style %%%%%%%%%%%%%%%%
\[
 \parbox{41mm}{ {\begin{fmfgraph}(42,10)
      \fmfleft{in}\fmfright{out}
      \fmf{brown_muck}{in,j}
      \fmf{vanilla}{j,jj}
      \fmf{vanilla,tension=3}{jj,v1}
      \fmf{vanilla}{v1,u}
       \fmf{vanilla,tension=2}{u,v2}
      \fmf{dashes}{v2,v3}
      \fmf{vanilla,tension=2}{v3,v4}
      \fmf{vanilla}{v4,v5}
      \fmf{brown_muck}{v5,out}
 \fmfv{decor.shape=circle,decor.filled=shaded,decor.size=4mm}{v1,u,v4}
    \end{fmfgraph}}}
\quad =:\frac{i{\tilde\Sigma}^2}{p^2-m^2-\Sigma}\,.
\]
\end{fmffile}

Taking all of these possibilities together we see that these
contribution to the propagator have the form
\begin{equation}
=\frac{i(1+\tilde\Sigma)^2}{p^2-m^2-\Sigma}\,.
\end{equation}

Of course  we can also  exchange  one or more  photons  from  one
dressing to the next.  This can be done for each and every one of
the above sets of diagrams.   We so obtain the following  rainbow
diagrams

\bigskip
 \begin{fmffile}{ufig44}
  %%%%%%%%%%%%%%%%% Defining brown_muck style %%%%%%%
\fmfcmd{%
vardef port (expr t, p) =
 (direction t of p rotated 90)
  / abs (direction t of p)
 enddef;}

\fmfcmd{%
  vardef portpath (expr a, b, p) =
   save l; numeric l; l = length p;
    for t=0 step 0.1 until l+0.05:
    if t>0: .. fi point t of p
      shifted ((a+b*sind(180t/l))*port(t,p))
     endfor
    if cycle p: .. cycle fi
 enddef;}

\fmfcmd{%
  style_def brown_muck expr p =
   shadedraw(portpath(thick/2,2thick,p)
    ..reverse(portpath(-thick/2,-2thick,p))
    ..cycle)
  enddef;}
%%%%%%%%%%%%%%%%% end of style %%%%%%%%%%%%%%%%
\[
\parbox{42mm}{\begin{fmfgraph*}(40,32)
      \fmfleft{in}\fmfright{out}
      \fmf{vanilla,tension=2}{in,j}
      \fmf{vanilla}{j,jj}
      \fmf{vanilla,tension=3}{jj,v1}
      \fmf{vanilla}{v1,u}
       \fmf{vanilla,tension=2}{u,v2}
      \fmf{dashes}{v2,v3}
      \fmf{vanilla,tension=2}{v3,v4}
      \fmf{vanilla}{v4,out}
      \fmfv{decor.shape=circle,decor.filled=shaded,decor.size=4mm}{v1,u,v4}
      \fmffreeze
      \fmf{photon,left=.82,
      lab=${  {  \cdot \atop n }\atop \cdot}$,
      lab.sid=right}{in,out}
       \fmf{photon,left=.40}{in,out}
    \end{fmfgraph*}}
+
     2\;\,\parbox{42mm}{\begin{fmfgraph*}(40,15)
      \fmfleft{in}\fmfright{out}
      \fmf{vanilla,tension=2}{in,j}
      \fmf{vanilla}{j,jj}
      \fmf{vanilla,tension=3}{jj,v1}
      \fmf{vanilla}{v1,u}
       \fmf{vanilla,tension=2}{u,v2}
      \fmf{dashes}{v2,v3}
      \fmf{vanilla,tension=2}{v3,v4}
      \fmf{vanilla}{v4,v5}
      \fmf{brown_muck}{v5,v6}
      \fmf{phantom,tension=9}{v6,out}
  \fmfv{decor.shape=circle,decor.filled=shaded,decor.size=4mm}{v1,u,v4}
       \fmffreeze
      \fmf{photon,left=.82,
      lab=${  { \cdot \atop n }\atop\cdot}$,
      lab.sid=right}{in,out}
       \fmf{photon,left=.40}{in,out}
    \end{fmfgraph*}}
+\;
  \parbox{42mm}{{\begin{fmfgraph*}(42,15)
      \fmfleft{in}\fmfright{out}
      \fmf{phantom,tension=9}{in,jjj}
      \fmf{brown_muck}{jjj,j}
      \fmf{vanilla}{j,jj}
      \fmf{vanilla,tension=3}{jj,v1}
      \fmf{vanilla}{v1,u}
       \fmf{vanilla,tension=2}{u,v2}
      \fmf{dashes}{v2,v3}
      \fmf{vanilla,tension=2}{v3,v4}
      \fmf{vanilla}{v4,v5}
      \fmf{brown_muck}{v5,v6}
      \fmf{phantom,tension=9}{v6,out}
    \fmfv{decor.shape=circle,decor.filled=shaded,decor.size=4mm}{v1,u,v4}
       \fmffreeze
      \fmf{photon,left=.82,
      lab=${  { \cdot \atop n }\atop\cdot}$,
      lab.sid=right}{in,out}
       \fmf{photon,left=.40}{in,out}
    \end{fmfgraph*}}}
\]
\end{fmffile}

\noindent  Using  the factorisation  property,  summing  over all
possible  numbers  of rainbow lines, and taking into account  the
$1/n!$  symmetry  factor  which  accompanies  a diagram  with $n$
rainbow lines, we can re-express this algebraically as
\begin{equation}\label{aop}
\frac{i(1+\tilde\Sigma)^2}{p^2-m^2-\Sigma}
\exp{\left(-C_{pp}\right)}
\,,
\end{equation}
where $-C_{pq}$ was defined in Eq.~\ref{Cdefn}.

Of course there can be other diagrams such as
\smallskip
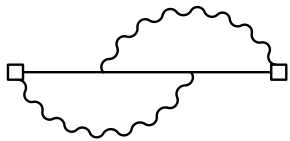
\begin{fmffile}{ufig45}
\[
  {\begin{fmfgraph}(35,13)
      \fmfleft{in}\fmfright{out}
      \fmf{vanilla}{in,j}
      \fmf{vanilla}{j,jj}
      \fmf{vanilla}{jj,out}
      \fmffreeze
      \fmf{photon,right=.7}{in,jj}
      \fmf{photon,left=.7}{j,out}
      \fmfv{decor.shape=square,decor.filled=empty,decor.size=2mm}{in}
      \fmfv{decor.shape=square,decor.filled=empty,decor.size=2mm}{out}
    \end{fmfgraph}}
\]
\end{fmffile}
\smallskip

\noindent where the dressing corrections overlap in a non-rainbow form, i.e.,
diagrams which are not one-particle  irreducible even after rainbow
lines are stripped off.  Such diagrams do not yield poles  and so
we can drop them here.   We neglect loops of matter fields since
these also remove the soft divergences.  Eq.~\ref{aop}\  is
therefore  our final result for the infra-red divergences  in and
the pole structure of the physical propagator.

We see  that  the dressings  factorise  into  a covariant  and  a
non-covariant  part.   For the full  non-covariant  wave function
renormalisation  constants,  we thus have  the attractive  result
that
\begin{equation}\label{aoztwo}
Z_2^p=Z_2^{\mathrm{cov}}(1+\tilde\Sigma)^2\exp(-C_{pp})
\,.
\end{equation}
Although   only   a  partial   exponentiation   of  the  dressing
contribution is apparent, we know that it is in fact IR-finite at
all orders. However, (\ref{aoztwo}) will suffice to show that the
pair creation vertex is finite to all orders.

\bigskip
\subsubsection[All Orders Pair Creation]{All Orders Pair Creation}
\smallskip
The general  class  of diagrams  with double  poles  and possible
IR-divergences in the pair creation process is

\medskip
 
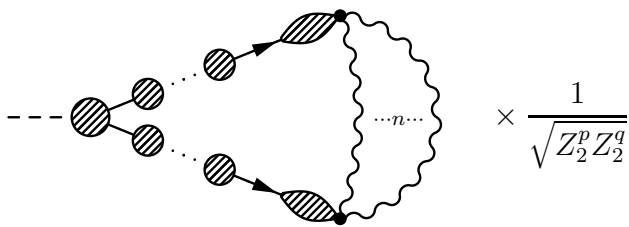
\begin{fmffile}{ufig46}
  %%%%%%%%%%%%%%%%% Defining brown_muck style %%%%%%%
\fmfcmd{%
vardef port (expr t, p) =
 (direction t of p rotated 90)
  / abs (direction t of p)
 enddef;}

\fmfcmd{%
  vardef portpath (expr a, b, p) =
   save l; numeric l; l = length p;
    for t=0 step 0.1 until l+0.05:
    if t>0: .. fi point t of p
      shifted ((a+b*sind(180t/l))*port(t,p))
     endfor
    if cycle p: .. cycle fi
 enddef;}

\fmfcmd{%
  style_def brown_muck expr p =
   shadedraw(portpath(thick/2,2thick,p)
    ..reverse(portpath(-thick/2,-2thick,p))
    ..cycle)
  enddef;}
%%%%%%%%%%%%%%%%% end of style %%%%%%%%%%%%%%%%
\[
  \parbox{50pt}{\begin{fmfgraph}(49,27)
      \fmfleft{source} \fmfright{pone,ptwo}
      \fmf{dashes,tension=0.7}{source,ver}
      \fmf{vanilla}{ver,v,v1}
      \fmf{dots,tension=0.8}{v1,v2,v3}
      \fmf{vanilla}{v3,v8}
      \fmf{fermion}{v8,uv}
      \fmf{vanilla,tension=5}{uv,vv}
      \fmf{brown_muck,tension=0.5}{vv,pone}
      \fmf{vanilla}{ver,u,v4}
      \fmf{dots,tension=0.8}{v4,v5,v6}
      \fmf{vanilla}{v6,v7}
      \fmf{fermion}{v7,vu}
       \fmf{vanilla,tension=5}{vu,v9}
      \fmf{brown_muck,tension=0.5}{v9,ptwo}
      \fmffreeze
      \fmfv{decor.shape=circle,decor.filled=shaded,decor.size=5mm}{ver}
      \fmfv{decor.shape=circle,decor.filled=shaded,decor.size=4mm}{v1}
      \fmfv{decor.shape=circle,decor.filled=shaded,decor.size=4mm}{v4}
      \fmfv{decor.shape=circle,decor.filled=shaded,decor.size=4mm}{v3}
      \fmfv{decor.shape=circle,decor.filled=shaded,decor.size=4mm}{v6}
      \fmfdot{pone,ptwo}
      \fmf{photon, right=0.95}{pone,ptwo}
      \fmf{photon, right=0.18}{pone,ptwo}
  \end{fmfgraph}}\qquad
  \qquad\qquad\qquad\!\!\!{\scriptstyle\cdots n\cdots}\qquad
    \times \frac1{\sqrt{Z_2^p Z_2^{q}}}
\]
\end{fmffile}

\noindent  where we have chosen  to multiply  the diagrams  by an
(IR-finite)  factor:   the inverse square root of the appropriate
wave   function   renormalisation    constant    (as   given   by
Eq.~\ref{aoztwo}) for each of the legs. This is useful as it makes
the exponentiation of the infra-red divergences apparent.

The  diagrams  we  have  retained   are  those  which  can  yield
IR-divergences,  i.e., where there are possible covariant  vertex
corrections from one leg to another, covariant corrections on the
external  legs, dressing  corrections  at the end of the external
legs and possible  rainbow corrections  from one dressing  to the
other.   The use of a black  blob, $\bullet$,  here denotes  that
there may or may not be a dressing correction ($\tilde\Sigma$) at
the ends of one or both of the lines.

Other diagrams  (e.g, where a line from a dressing connects  to a
covariant interaction vertex on the other line) will not give two
poles or will be infra-red finite.

We now factorise the rainbow dressings,  this means (summing over
all possible  rainbow  lines  and including  the $1/n!$  symmetry
factor for a diagram with $n$ such lines) that we may write

 \begin{fmffile}{ufig47}
  %%%%%%%%%%%%%%%%% Defining brown_muck style %%%%%%%
\fmfcmd{%
vardef port (expr t, p) =
 (direction t of p rotated 90)
  / abs (direction t of p)
 enddef;}

\fmfcmd{%
  vardef portpath (expr a, b, p) =
   save l; numeric l; l = length p;
    for t=0 step 0.1 until l+0.05:
    if t>0: .. fi point t of p
      shifted ((a+b*sind(180t/l))*port(t,p))
     endfor
    if cycle p: .. cycle fi
 enddef;}

\fmfcmd{%
  style_def brown_muck expr p =
   shadedraw(portpath(thick/2,2thick,p)
    ..reverse(portpath(-thick/2,-2thick,p))
    ..cycle)
  enddef;}
%%%%%%%%%%%%%%%%% end of style %%%%%%%%%%%%%%%%
\[
  \parbox{44pt}{\begin{fmfgraph}(43,27)
      \fmfleft{source} \fmfright{pone,ptwo}
      \fmf{dashes,tension=0.7}{source,ver}
      \fmf{vanilla}{ver,v,v1}
      \fmf{dots,tension=0.8}{v1,v2,v3}
      \fmf{vanilla}{v3,v8}
      \fmf{fermion}{v8,uv}
      \fmf{vanilla,tension=5}{uv,vv}
      \fmf{brown_muck,tension=0.5}{vv,pone}
      \fmf{vanilla}{ver,u,v4}
      \fmf{dots,tension=0.8}{v4,v5,v6}
      \fmf{vanilla}{v6,v7}
      \fmf{fermion}{v7,vu}
       \fmf{vanilla,tension=5}{vu,v9}
      \fmf{brown_muck,tension=0.5}{v9,ptwo}
      \fmffreeze
      \fmfv{decor.shape=circle,decor.filled=shaded,decor.size=5mm}{ver}
      \fmfv{decor.shape=circle,decor.filled=shaded,decor.size=4mm}{v1}
      \fmfv{decor.shape=circle,decor.filled=shaded,decor.size=4mm}{v4}
      \fmfv{decor.shape=circle,decor.filled=shaded,decor.size=4mm}{v3}
      \fmfv{decor.shape=circle,decor.filled=shaded,decor.size=4mm}{v6}
      \fmfdot{pone,ptwo}
    \end{fmfgraph}}\qquad \qquad
    \times
         \exp\left(-C_{pq}\right)
          \frac1{\sqrt{Z_2^pZ_2^{q}}}
\]
\end{fmffile}

\smallskip
\noindent Since there may or may not be a dressing correction  at
the  end  of these  lines,  we may  write  these  end factors  as
$(1+\tilde\Sigma)$.    They  clearly  then  \textit{cancel}   the
$\tilde\Sigma$  dependence  from  each  of the external  leg wave
function  renormalisation  factors  as  given  by (\ref{aoztwo}).
Diagrammatically we thus have

\medskip

 \begin{fmffile}{ufig48}
\[
  \parbox{40pt}{\begin{fmfgraph}(38,25)
      \fmfleft{source} \fmfright{pone,ptwo}
      \fmf{dashes,tension=0.8}{source,ver}
      \fmf{vanilla}{ver,v,v1}
      \fmf{dots}{v1,v2,v3}
      \fmf{vanilla}{v3,v8}
      \fmf{fermion}{v8,vv}
      \fmf{vanilla,tension=1.5}{vv,pone}
      \fmf{vanilla}{ver,u,v4}
      \fmf{dots}{v4,v5,v6}
      \fmf{vanilla}{v6,v7}
      \fmf{fermion}{v7,v9}
      \fmf{vanilla,tension=1.5}{v9,ptwo}
      \fmffreeze
      \fmfv{decor.shape=circle,decor.filled=shaded,decor.size=5mm}{ver}
      \fmfv{decor.shape=circle,decor.filled=shaded,decor.size=4mm}{v1}
      \fmfv{decor.shape=circle,decor.filled=shaded,decor.size=4mm}{v4}
      \fmfv{decor.shape=circle,decor.filled=shaded,decor.size=4mm}{v3}
      \fmfv{decor.shape=circle,decor.filled=shaded,decor.size=4mm}{v6}
    \end{fmfgraph}}\qquad \qquad
    \times \exp\left( {-C_{pq}+\frac12 C_{pp}+ \frac12 C_{qq}}\right)
         \frac1{Z_2^{\mathrm{cov}}}
\]
\end{fmffile}

\noindent   and   we  see  that   the   dressing   effects   have
exponentiated.   Since the covariant diagrams which are left over
are well known to exponentiate,  we see that \textit{all the soft
and  phase  effects  exponentiate}  in the residue  of the double
poles.   We have  seen  that they cancel  at one loop,  this  now
trivially  implies that \textit{the  dressed process is IR-finite
at all orders at the level of Green's functions}.  We stress that
this holds for both soft and phase divergences.  The extension of
this  approach   to  scattering   and  to  higher   vertices   is
straightforward.

\bigskip

We have thus seen that the dressings  we have  introduced in
Sect.~2 remove soft and phase divergences in massive
Electrodynamics already at the   level   of  Green's   functions
and  permit   a particle interpretation. The immediate  question
now is whether these perturbative successes can be extended  to
QCD.  In particular, we need to clarify  the generalisation  to  a
different  class  of mass singularities, collinear  divergences,
which  arise  in QCD. Since collinear divergences   also
characterise   QED with massless charged particles,  the  next
section  is dedicated  to a study of that theory.

\section[Massless Charges]{Massless Charges}

Massless charges are an essential ingredient of the standard model.
In QCD the colour charge of the massless gluon results in its
self-interaction which in turn leads to asymptotic freedom. In
addition, gluonic bremsstrahlung plays a dominant role in jet
formation~\cite{Dokshitser:1991wu}. However, such massless charges
are poorly understood, even at the level of cross-sections, since
the Block-Nordsieck procedure of summing over degenerate final
states fails to yield finite results~\cite{Doria:1980ak,
Di'Lieto:1981dt}. Massless charges can be modelled in QED by taking
the limit of vanishing electron mass. As one would expect, QED then
has many new and unusual features. In particular, massless electron
pairs can now be created by expending arbitrarily small amounts of
energy. This freedom leads to a new type of singularity, the
collinear divergences. In perturbation theory these express
themselves through the divergence of on-shell massless propagators.
The inverse propagator, $(p-k)^2-m^2$, vanishes in the $m\to 0$
limit if $p$ is on-shell and $k$ is parallel to $p$. (Note that
$\kb$ is not necessarily small.) For further details
see~\cite{weinberg:1995}. The matter pairs, which can now be easily
created, screen the initial charge and the effective coupling
vanishes~\cite{Espriu:1996sk}. Here we will see that the argument
of Sect.\ 2 for the vanishing of the asymptotic interaction
Hamiltonian, if correctly dressed matter is used, also applies in
the massless limit. Finally we discuss some aspects of the
construction of dressings in this limit.

\bigskip
\subsection{Collinear Asymptotic Dynamics}
\smallskip

The asymptotic behaviour of the interaction Hamiltonian for massive
QED was discussed in Sect.\ 2. In the massless case we will show
that the interaction Hamiltonian in the distant past or future has
a far richer structure than in the massive case.

The interaction Hamiltonian, $\Ha(t)$, is given by (\ref{intham}),
where the conserved matter current and the free field expansions
for matter and gauge fields are still those of (\ref{psi_free}) and
(\ref{a_free}). The all-important difference is that, since $m=0$,
the energy in the massless case is given by $E_p=|\pb |$.

When the expansions (\ref{psi_free}) and (\ref{a_free}) are
substituted into (\ref{intham}), then, of the eight possible terms,
the two involving $a_\mu \overline{v} \gamma ^\mu u$ and
$a^\dagger_\mu
\overline{u} \gamma^\mu v$ will have a time dependence
of the form $e^{i\alpha t}$ with $\alpha$ being either positive or
negative. These cannot survive for large values of $t$ and so do
not appear in the asymptotic limit. This leaves six terms which are
potential survivors, of which two vanish in the massive case. A
closer look at some of these structures will show how to evaluate
all the terms.

The first term to be considered in detail is one of the \lq\lq
off-diagonal\rq\rq\ (in the spinors) terms:
\begin{equation}
\label{hterm1}
-\ee\int \frac{d^3x\,d^3k\,d^3p\,d^3q}{(2\pi)^9}
\frac{a_\mu(k)}{2\omega_k
\sqrt{4E_pE_{q}}}
b^\dagger (q,s)d^\dagger(p,r)\overline{u}^s(q)\gamma^\mu
v^r(p)e^{-ik\cdot x} e^{iq\cdot x}e^{ip\cdot x}\,.
\end{equation}
Integrating out the $\boldsymbol{x}$-integral will give a term
involving $\delta(\qb +\pb -\kb )$. Following this by integrating
out the $\pb $ integral, the expression (\ref{hterm1}) becomes
\begin{equation}
\label{hterm2}
-\ee\int \frac{d^3k\,d^3q}{(2\pi)^6}\frac{a_\mu(k)}{2\omega_k
\sqrt{4E_qE_{k-q}}}b^\dagger (q,s)d^\dagger (k-q,r)
\overline{u}^s(q)\gamma^\mu v^r(k-q)e^{it(E_q+E_{k-q}-\omega_k)}\,.
\end{equation}
\no If this is to survive as $t\rightarrow \infty$ then the coefficient
in the exponent must vanish. This is equivalent to demanding that
$\omega_k=E_q+E_{k-q}$, or
$|\kb |=|\qb |+|\boldsymbol{
k}-\qb |$. Since these three vectors represent the three
sides of a triangle, plainly a solution to this equation will exist
if and only if the vectors $\kb $ and $\qb $
are parallel and $|\kb |\geq |\qb |$. In this
region, where the photon is
\emph{collinear} with the matter field, this contribution to the
asymptotic interaction Hamiltonian does not vanish. The momentum,
$\kb $, is no longer restricted to the soft region, $\kb =0$, but
must be larger than the matter momenta, since this term corresponds
to pair creation.

 The second term to be examined is one of the \lq\lq
diagonal\rq\rq\  terms and has the form
\begin{equation}
\label{hterm3}
-\ee\int \frac{d^3x\,d^3k\,d^3p\,d^3q}{(2\pi)^9}
\frac{a_\mu(k)}{2\omega_k
\sqrt{4E_pE_{q}}}
d(q,s)d^\dagger(p,r)\overline{v}^s(q)\gamma^\mu v^r(p)e^{-ik\cdot
x} e^{-iq\cdot x}e^{ip\cdot x}\,.
\end {equation}
Integrating out the $\boldsymbol{x}$ again gives a delta function,
this time $\delta(\pb -\kb -\qb )$. Integrating out the $\qb $
integral now yields
\begin{equation}
\label{hterm4}
-\ee\int \frac{d^3k\,d^3p}{(2\pi)^6}\frac{a_\mu(k)}{2\omega_k
\sqrt{4E_pE_{p-k}}}d(p-k,s) d^\dagger (p,r)
\overline{v}^s(p-k)\gamma^\mu v^r(p)e^{it(E_p-E_{p-k}-\omega_k)}\,.
\end{equation}
The exponent in (\ref{hterm4}) must vanish for large $t$ and this
implies that $E_p-E_{p-k}-\omega_k$ must vanish. This is similar to
the previous case except that now we must have
$|\pb |=|\kb |+|\pb -\kb |$.
Reasoning as before, we find that $\pb $ must be parallel
to $\kb $ but now $|\pb |
\geq |\kb |$, i.e., $E_p \geq \omega
_k$. This clearly corresponds to photon production.

 If the six terms are evaluated
using  (\ref{terms2}) then the final form of the asymptotic
interaction Hamiltonian in the massless case
is~\cite{havemann:1985}
\begin{equation}
\label{hasymp}
\Ha^{\mathrm{as}}(t)=-\ee\int\frac{d^3k}{2\omega_k}[a_\mu(k)J^\mu_{as}(k,t)
+\mathrm{h.c.}]\,,
\end{equation}
where \lq h.c.\rq\  denotes the hermitian conjugate of the first
term in brackets, and
\begin{equation}
\label{current}
J^\mu_\mathrm{as}(k,t)=\int\frac{d^3p}{(2\pi)^3}\frac{p^\mu}{E_p}
\Bigl(\rho_\mathrm{scatt}(p,k) +\rho_\mathrm{prod}(p,k)\Bigr)\,.
\end{equation}
The two structures in the asymptotic current are respectively
\begin{equation}
\label{scatt}
\rho_\mathrm{scatt}(p,k)=\sum_s[b^\dagger(p,s)b(p-k,s)-d^\dagger(p,s)d(p-k,s)]
e^{it(E_p-E_{p-k}-\omega_k)}
\end{equation}
in the region $\omega_k\leq E_p$, and
\begin{equation}
\label{prod}
\rho_\mathrm{prod}(p,k)=\sum_sb^\dagger
(p,s)d^\dagger(k-p,r)\xi^{s\dagger}\boldsymbol{\sigma}
\cdot\hat{\boldsymbol{
q}}\xi^re^{it(E_p+E_{p+k}-\omega_k)}
\end{equation}
in the region $\omega_k\geq E_p$. The two terms in (\ref{hasymp})
have different physical interpretations: the $\rho_\mathrm{scatt}$
part corresponds to photon radiation and includes the soft region
which was responsible for the non-vanishing of the usual soft
asymptotic interaction Hamiltonian,~(\ref{hint_as}). However, the
photon momenta is only required to be collinear and not extremely
soft. The other term, $\rho_\mathrm{prod}$, is completely new and
corresponds to the production of massless matter pairs with momenta
less than that of the initial photon. Even though photons cannot
radiate other photons, these two structures in this model theory
already show the basic processes underlying the collinear
production of gluons and quark-antiquark pairs in jet creation.

We now note that from the form of~(\ref{hasymp}), and in particular
the explicit $p^\mu$ factor in the asymptotic current, the solution
to the dressing equation, (\ref{dress1}) or (\ref{dresseqtn}),
will,  also here in the massless theory, correspond to a particle
since the asymptotic interaction Hamiltonian vanishes at the
correct point on the mass shell. We thus predict that
\emph{the Green's functions of the dressed fields will also be free of
collinear divergences}.

In the context of this discussion of the asymptotic interaction
Hamiltonian, it should be noted that, for massive electrons, if we
allow the photon a small mass there is no momentum configuration
such that the exponential survives at asymptotic times. In
perturbation theory it may be easily seen that a small photon mass
regulates the infra-red singularities. However, if the mass of the
photon becomes large enough to open decay channels into matter
pairs, then once again the interaction picture breaks down and this
results in apparently gauge dependent S-matrix elements (see,
e.g.,~\cite{Nowakowski:1993iu}).

We now turn to the construction of the solutions of the dressing
equation in the massless limit.

\bigskip
\subsection{Collinear Dressings}
\smallskip
In the massless theory, the dressing must still fulfill the
requirements of gauge invariance~(\ref{h_trans}) and the dressing
equation~(\ref{dresseqtn}), where now $(\eta+v)^2=0$. Following the
procedure outlined in Sect.~2 for solving these equations, we will
require the commutators of the potential in the theory described by
the asymptotic current, (\ref{current}). As we saw for the massive
case, the potential can now be written as in (\ref{sola}). However,
in this massless theory, the asymptotic currents no longer commute
and the commutators of the theory are no longer those of the free
theory. This makes the construction of the dressing of massless
charges somewhat more difficult. Rather than discuss the details of
how to extend the analysis of Sect.~2, we will now examine the
$v\to 1$ limit of the massive case and then study, in this limit,
the route taken by Dirac to the dressings.

The naive massless limit of a dressed charge~(\ref{hexponentiated})
moving in the $x^1$-direction results in integrals of the form
\begin{equation}\label{naive}
-\frac{1}{4\pi}\int d^3z
\frac{(\partial
_2 A_2 +\partial
_3A_3-E_1)(x^0,\boldsymbol{z})}{|x^1-z^1|}\,.
\end{equation}
Clearly this is ill-defined, and although $1/|x^1-z^1|$ can be
defined in terms of generalised functions~\cite{jones:1966}, the
serious problems with this naive limit cannot be circumvented in
this fashion. Further evidence that a naive approach to the
massless theory will not suffice, comes from the detailed
perturbative calculation of the one-loop
propagator~\cite{Bagan:1997dh, Bagan:1997su}. The wave-function
renormalisation constants found in these papers diverge as $v\to1$.

To obtain a deeper insight into the form of the dressings, we can
follow Dirac's lead and construct a dressing that reproduces the
electric and magnetic fields for a massless charge. The electric
field of an ultra-relativistic charge contracts in the direction of
motion. For a massless charge this contraction leads to a singular
field configuration. It has been argued that, for motion in the
$x^1$-direction, the electromagnetic fields are~\cite{kabat:1992,
robinson:1984}
\begin{eqnarray}
\label{field}
E_1=0,& & E_i=\frac{-ex^i\delta(x^0-x^1)}{2\pi r^2}\nonumber \\
& & \\
 B_1=0,& &B_i=\frac{\varepsilon^{ij}x^j\delta(x^0-x^1)}{2\pi r^2}
 \nonumber
\end{eqnarray}
where $i,j=2,3$ and $r^2=(x^2)^2+(x^3)^2 = |x_\perp |^2$. This
result should follow from the soft dressing term.

We argue that this part of the dressing should have the form
\begin{equation}
\label{green2}
\chi _1(x^0,\boldsymbol{x}) =\frac{1}{2\pi}\int d^2z^\perp (\partial _2 A_2 +
\partial _3A_3-E_1)(x^0,x^1,z^\perp)\log |x^\perp -z^\perp | + g(x)
\end{equation}
where $g(x)$ is a function which is harmonic in $x^\perp$, i.e.,
$g(x)$ is in the kernel of the two dimensional Laplacian. Although
this is not the naive limit (\ref{naive}), it can be understood as
arising from that part of the dressing equation which describes the
soft dynamics, $\G\cd \pa( \chi) = \G\cd A$. For this
configuration, the equation becomes
\begin{equation}
\label{green1}
((1-v^2)\partial _1^2 +\partial _2^2  +\partial _3^2)\chi
_v(x^0,\boldsymbol{x})=((1-v^2)\partial
_1A_1+\partial _2 A_2 +\partial _3A_3-vE_1)(x^0,\boldsymbol{x})\,.
\end{equation}
Now in the $v\to1$ limit, one sees that the limiting value of
$\chi_v$, which we write $\chi_1$, must satisfy
\begin{equation}
\label{light}
(\partial _2^2  +\partial _3^2)\chi
_1(x^0,\boldsymbol{x})=
(\partial _2 A_2 +\partial _3A_3-E_1)(x^0,\boldsymbol{x})\,,
\end{equation}
with solution (\ref{green2}). It may be checked that the dressing
constructed out of this soft term, with $g(x)$ set to zero, indeed
yields the electric and magnetic fields, (\ref{field}). This result
strengthens our claim that a careful analysis of the dressing
equation will allow us to construct massless charges.

\section[Discussion]{Discussion}

%\input{sec6}

%%%%%%%%%%%%%%%%%%%%%%%%%%%%%%%%%%%%%%%%%%%%%%%%%%%%%%%%%%%%%%%%
%% Sect. 6 of the Curry of Quarks
%%
%% Started 19/3/1998  ML
%%
%%
%%%%%%%%%%%%%%%%%%%%%%%%%%%%%%%%%%%%%%%%%%%%%%%%%%%%%%%%%%%%%%%%

%% The section

\noindent In this review we have demonstrated that it is
indeed possible to construct relativistic charged particles.
Let us now recall what we have seen.

Our starting point was the non-vanishing of the asymptotic
interaction Hamiltonian which characterises gauge theories. The
misidentification of the asymptotic interactions of the free
Hamiltonian with the true asymptotic dynamics of gauge theories
causes the infra-red problem. Since the interaction Hamiltonian
does not vanish at large times, Gauss' law tells us that charged
particles are not just the matter fields of the Lagrangian:
physical particles like the electron are always accompanied by an
electromagnetic dressing. In practical calculations this expresses
itself in the lack of a pole structure in the on-shell Green's
functions of the Lagrangian matter fields and in divergences in
S-matrix elements.

Any physical degree of freedom must be gauge invariant. For charged
particles this means that the dressing together with the matter
field at its core must be locally gauge invariant --- this
translates into an equation for the dressing, (\ref{h_trans}). This
requirement, though, does not suffice to construct charged
particles: any gauge invariant solution is in principle a physical
degree of freedom, but it is not necessarily one that physics
chooses to use. To restrict ourselves to the solutions which are
physically relevant, we require a second equation,
(\ref{dresseqtn}). This latter relation was deduced from demanding
that the velocity of an incoming or outgoing charged particle is
well defined and the dressing must take this into account. The form
of the asymptotic interaction Hamiltonian is such that it vanishes
in the propagator of a correctly dressed particle, i.e., one which
satisfies (\ref{h_trans}) and (\ref{dresseqtn}). It follows that
dressed charges are then described by the free Hamiltonian and a
relativistic particle description is indeed possible. In Sect.\ 2
dressings which solve these equations were explicitly constructed.
These are physical degrees of freedom with a specific physical
interpretation. The dressings factorise into two parts, each of
which was interpreted as playing a different role in the infra-red
physics of QED. Sect.\ 3 then illuminated the role of velocity in
any description of charged particles and showed how the form of the
dressing could be derived from the theory of heavy charges.

These arguments have been tested by explicit perturbative
calculations described in Sect.\ 4. The interpretations of the two
different terms in the dressing were such that they were expected
to each cancel a different type of infra-red divergence. The
constructions passed these one-loop tests with flying colours and a
gauge invariant pole structure was obtained. This was then
generalised to an all orders proof of the cancellation of infra-red
divergences in the on-shell Green's functions of QED with dressed
fields. These results give us great confidence in the
requirements~(\ref{h_trans}) and (\ref{dresseqtn}), in the
solutions~(\ref{hexponentiated}) and also in the interpretation we
associate with these physical degrees of freedom.

Finally in Sect.\ 5 collinear divergences were studied in the
framework of QED with massless fermions. Here the structure of the
interaction Hamiltonian which survives at large times is much
richer than in QED with massive matter. However, it still vanishes
if correctly dressed matter, satisfying~(\ref{dresseqtn}), is used.
This implies that the on-shell Green's functions of the solutions
to (\ref{h_trans}) and (\ref{dresseqtn}) in the massless theory
will also be free of all infra-red singularities, including
collinear divergences. The construction of these dressings was
considered.

\subsection[Charges in the Standard Model]{Charges in the Standard Model}
\smallskip
Most of this paper has been given over to QED. The other
interactions in the standard model of particle physics are also
described by gauge theories, however, the charged particles
in these theories are very different to those of the unbroken
abelian theory. In this subsection we will sketch some of the
most important differences which characterise the charged particles
of the weak and strong nuclear forces.

The asymptotic Hamiltonian in QCD also does not reduce to the free
one. However, there is a new problem with defining charges in
non-abelian gauge theories such as QCD. The colour charge
\begin{equation}
Q^a=\int d^3x(J^a_0(x)-f^a_{bc}E^b_i(x)A^c_i(x))
\,,
\end{equation}
in
sharp contrast to that of QED, is \emph{not} invariant under
gauge transformations. It is therefore natural to wonder how we
may speak of colour charged particles. But on physical states,
where Gauss' law holds, the charge may be written as
\begin{equation}\label{Qwho}
Q^a=\frac1{g}\int d^3x\partial_i E_i^a(x)
\,.
\end{equation}
Using Gauss' theorem
it follows that the colour charge expressed in this way is in
fact invariant under gauge transformations which reduce, in a
directionally independent manner, to
elements of the centre of the group at spatial infinity. Thus the
concept of colour charge is only meaningful for locally gauge
invariant fields and then only if this restriction on the allowed gauge
transformations is imposed. Since the Lagrangian
matter fields are not gauge
invariant, coloured quarks in QCD are necessarily dressed
by glue~\cite{Lavelle:1996tz, Lavelle:1997ty}.

The dressing equation~(\ref{dresseqtn}) of QED may be directly
generalised~\cite{Bagan:1998kg} to QCD. The non-abelian solutions
will be asymptotically described by a free Hamiltonian provided
that the field transformations required are, in fact, admissible.
To the extent that they may be constructed, it follows that the
various mass singularities would cancel at the correct points in
the mass shell of the dressed quark fields.

However, there is a fundamental obstruction to the construction of
dressed quarks (and gluons) which follows from the nature of
non-abelian gauge transformations. Any gauge invariant description
of a quark could be used to construct a gauge fixing. However, the
boundary conditions which must be imposed on the allowed gauge
transformations in order for colour to be associated with the
physical degrees of freedom, are such that the Gribov
ambiguity~\cite{Gribov:1978wm, Singer:1978dk} holds. There is then
a fundamental, non-perturbative limit on the construction of
gauge-invariant, coloured charges. Thus \emph{the true degrees of
freedom in QCD outside of the perturbative domain do not include
quarks and gluons}. Colourless gauge-invariant fields can of course
be constructed, but quarks and gluons are confined.

The picture of confinement
which emerges from studying the construction of coloured charges
is as follows. When a $Q\bar Q$-system is
separating, but the matter fields are still
at a short distance from each other, the interaction potential
is essentially Coulombic and the dynamics is described by
perturbative dynamics. The short-distance, Coulombic inter-quark potential
may be described using low-order perturbative solutions to the
dressing equations for each of the individual quarks. Deviations from
this such that a confining potential arises in
a $Q\bar Q$-system are expected to come from a \lq mesonic dressing\rq\
which does not factorise into two parts.

Finally, it might be objected that the non-abelian theories which
underly the weak interaction do not lead to confinement of weakly
charged particles such as the $W$ or indeed the electron. This is
easily understood: in spontaneously broken gauge theories we are
entitled to use the Higgs sector to dress charged
particles~\cite{Lavelle:1995rh}. (The ability to choose the unitary
gauge circumvents the gauge fixing ambiguity.) In this way gauge
invariant solutions corresponding to weakly charged particles may
be constructed.

\subsection[Open Questions]{Open Questions}
\smallskip
This review has sketched out a systematic approach to the construction of
charged particles. The qualitative results to date of the program are fully in
accord with phenomenology and the calculational tests to which the
charged particles have been subjected have all supported the
methods. But there are still a large number of unanswered questions
ranging from calculational procedures to the extension of the applicability of
this programme to physics at finite temperature and density.
We will now conclude by listing some of the most pressing tasks.

\medskip

\no \textbf{a)} The above proof of the infra-red finiteness
of the dressed on-shell Green's functions needs to be
extended to full calculations of these Green's functions, in
particular the UV renormalisation of the $n$-point functions of these
composite operators must be performed.
The dressed propagators of both fermionic and scalar QED have been
carried out at one-loop. This now needs to be extended to higher
loops and to vertex functions.

\no\textbf{b)} The perturbative tests need to be extended to
collinear divergences. Massless QED is the natural
testing ground here. First the dressing equation needs to be
solved within the framework of the remarks of Sect.\ 5. Then the
solutions of the equivalent dressing equation for QCD should be
constructed and tested.

\no\textbf{c)} Although a brute force, direct perturbative
solution of the dressing equations for the non-abelian theory is
feasible at low orders, a systematic and practical approach to the
construction of a gauge invariant dressed quark fulfilling the
dressing equation of QCD for an arbitrary velocity is needed. In
the appendix to Ref.~\cite{Lavelle:1997ty} an algorithm was
presented with which the calculation of a particular gauge
invariant dressed quark solution to any order in perturbation
theory became rather simple and some of the first terms in this
perturbative construction were given. In this context we also
refer to Ref.~\cite{haller:1997} where all orders expressions
apparently corresponding to the gauge invariant extension of this
term were presented. This corresponded to the gauge invariant
extension of the soft term in the QED dressing of a static charge.
Such work needs to be generalised to both terms and indeed to
arbitrary velocities. Furthermore the success of phenomenological
constituent quark models in describing hadronic structure strongly
indicates that, within the overall constraint of confinement, some
non-perturbative input into the construction of dressed quarks
should be possible. Here we are thinking especially of the role of
instantons and condensates in chiral symmetry breaking. We stress
that it will not be possible to incorporate all the
non-perturbative, topological aspects of QCD into dressings of
individual quarks or gluons. This, we have argued above, is how
confinement makes itself manifest.

\no\textbf{d)} The importance of the  perturbative
chromo-electric and chromo-magnetic dressings is that they
determine the short distance
interactions between quarks and have implications for jet physics.
We would urge, e.g., a comparative study of the distribution of
glue in the dressings around quarks and gluons which could shed
light on the different development of such jets. The fruitful
concept of parton-hadron duality~\cite{Khoze:1997dn} could we suggest
be potentially
replaced by a refined version of a duality between (perturbatively)
gauge invariant, dressed colour charges and the resulting physical
hadrons.

\no\textbf{e)} The qualitative proof of quark confinement
which was described above needs to be quantified. How can
the hadronic scale be found in this way?
That quarks are not part of the
physical degrees of freedom of QCD will only become apparent at
larger distances where that part of the non-perturbative dynamics
which is sensitive to the Gribov ambiguity becomes first
significant and finally dominant. This is, of course, a tough
non-perturbative calculation. One natural technique is the
lattice, where there has been some work on studying the gauge
fixing problem. It will be important here to distinguish between
lattice artifacts and the true, physical limitation on gauge
fixing~\cite{Parrinello:1991fp, deForcrand:1995mz}.
We also suggest here that the construction of dressed charges
in monopole backgrounds be studied.

\no\textbf{f)} In view of the difficulties inherent in
non-perturbative calculations, phenomenological modelling of
dressings is desirable. Thus we feel that the construction of
dressed quarks with phenomenologically desirable properties, e.g.,
running masses should be investigated.
A corollary of this last point is the construction
of mesonic dressings which lead to phenomenological
inter-quark potentials. The stability of such model dressings
should then be tested using variational methods. We recall from
the introduction the instability of the (confined)
\lq$e^+e^-$\rq -system where the two matter fields are linked by a
string~\cite{Prokhorov:1993, Haagensen:1997pi}.

\no\textbf{g)} One of the most important questions in particle physics
is how mass is generated. The construction of charges in theories
with spontaneous symmetry breaking deserves further, quantitative
investigation. Both perturbative and non-perturbative effects
should be studied. A major question closely related to this is how
one should describe unstable charged particles, such as those
occurring in the weak interaction. It is well known that there are
problems with obtaining gauge invariant results in the presence of
such fields~\cite{Nowakowski:1993iu, Stuart:1995ba,
Papavassiliou:1995fq}. We have seen, at the end of  Sect.~5.1,  the
non-vanishing of the asymptotic interaction Hamiltonian in such
theories. We urge the development of, in some sense dressed,
admixtures of fields which would fulfill the following properties:
i) if the coupling is set by hand to zero it should reduce to the
Lagrangian field, \linebreak ii) they must be gauge invariant and
iii) the asymptotic interaction Hamiltonian should vanish for the
propagator of these constructs.

\no\textbf{h)} To what extent can we talk about charged particles
beyond the standard model? The construction of descriptions of
charged particles in theories such as technicolour and
with (broken) supersymmetry could cast light on their experimental
signatures. Here it is important to see if the
gauge invariant, dressed charged particles are still related by
supersymmetries, or if this just holds for the Lagrangian fields.
Furthermore we strongly suggest a study of the construction and dynamics of
charged particles in unbroken supersymmetric models so as to
clarify the role of (supersymmetrically) charged particles in the
Seiberg-Witten description of confinement in such theories~\cite{Seiberg:1994rs}.

\no\textbf{i)} Constrained dynamics is the mathematical framework
for extracting physical degrees of freedom in theories such as QED
and QCD. In modern formulations BRST symmetry is used to single out
gauge invariant, local fields. However, as will have become
apparent above, gauge invariance is not enough: one must still make
a clear identification between the true degrees of freedom and the
observed particles.  Quark confinement in QCD is merely the most
obvious example; even in Quantum Electrodynamics the infrared
problem, and the associated superselection structures labelling
charges with different velocities, show that this identification is
not direct. The interplay between the asymptotic interaction
Hamiltonian and the physical observables needs further study. It
should be noted that the BRST method is a local construction and,
as we have seen, charged fields are necessarily non-local. In this
context we point out a further symmetry of QED which was noted in
Ref.~\cite{Lavelle:1993xf} and which singled out a subset of gauge
invariant fields. We believe that this is only one representative
of a class of symmetries with whose aid the true degrees of freedom
may be interpreted. Finally, the incorporation of nonperturbative
effects is the outstanding question in mathematical physics today.
In constrained dynamics this entails extending the usual reduction
procedure to theories with superselection sectors characterised by
the existence of nontrivial surface terms and other global
structures. A proper understanding of these effects is essential if
constrained dynamics is to be able to discuss nonperturbative
physics whether it be the quark structure of low energy QCD or
string phenomenology.

\bigskip\medskip
\no\textbf{Acknowledgements:} First and foremost we wish to thank
Emili Bagan in collaboration with whom the work of Sect.~4 was
carried out. Between us we also thank the following for
discussions and correspondence: David Broadhurst, Emilio d'Emilio,
Tomeu Fiol, O.W.\ Greenberg, Kurt Haller, Marek Nowakowski, John
Ralston, Nicolas Roy, Dieter Sch\"utte, Sergei Shabanov, Manfred
Stingl, Shogo Tanimura, Rolf Tarrach, John C. Taylor and Izumi
Tsutsui.

\appendix

\section[Appendix]{Appendix}

To obtain the asymptotic Hamiltonians of Sect.\ 2 and Sect.\ 5 we
must evaluate the terms $\overline{u}^r(p)\gamma^\mu u^s(q)\,$,
$\overline{v}^r(p)\gamma^\mu v^s(q)\,$,
$\overline{v}^r(p)\gamma^\mu u^s(q)$ and $
\overline{u}^r(p)\gamma^\mu v^s(q)$.
The results we will obtain, and indeed some generalisations, are
quoted in Appendix J of Ref.~\cite{Bialynicki-Birula:1975yp}.

\no Given an on-shell $4$-vector, $p$, so that $p^2=m^2$, then $E_p$
is given by $E_p=\sqrt{|\pb |^2+m^2}$. We introduce the notation
\begin{equation}
\label{not1}
N_p=\frac{1}{E_p+m} \qquad \mbox{and} \qquad
\hat{\pb }=N_p\pb  \,.
\end{equation}
Let $\xi^{1\dagger}=(1,0)$ and $\xi^{2\dagger}=(0,1)$. The Dirac
spinors for $r=1,2$ are  taken as
\begin{equation}
u^r(p)= \frac{1}{\sqrt{N_p}}
\left(
\begin{array}{c}
\xi^r \\
\boldsymbol{\sigma }\cdot \hat{\pb } \xi^r
\end{array}
\right)
\, \qquad
v^r(p)= \frac{1}{\sqrt{N_p}}
\left(
\begin{array}{c}
\boldsymbol{\sigma}\cdot \hat{\pb } \xi^r\\
\xi^r
\end{array}
\right)
\end{equation}
The $u^r(p)$ and $v^r(p)$ are, respectively, positive and negative
energy solutions to the Dirac equation. Let us define
\begin{equation}
\Lambda=\left(
\begin{array}{cc}
0&1 \\ 1&0
\end{array}
\right)\,.
\end{equation}
Then we have
\begin{equation}
u^r(p)=
\Lambda
v^r(p)\,,
\end{equation}
and
\begin{equation}
\overline{u}^r(p)=
\overline{v}^r(p)\gamma ^0
\Lambda
\gamma ^0\,.
\end{equation}
Our gamma matrix convention is
\begin{equation}
\gamma^0=
\left(
\begin{array}{cc}
1&0 \\ 0&-1
\end{array}
\right)
\,, \qquad
\gamma ^i=
\left(
\begin{array}{cc}
0&\sigma^i \\ -\sigma^i&0
\end{array}
\right)\,,
\end{equation}
where $\sigma^i$ are the Pauli matrices. It is now straightforward
to show that
\begin{eqnarray}
\overline{u}^r(p)\gamma^\mu u^s(q)=&\overline{v}^r(p)\gamma^\mu
v^s(q)\,,\\
\overline{u}^r(p)\gamma^\mu v^s(q)=&\overline{v}^r(p)\gamma^\mu u^s(q)\,.
\end{eqnarray}
This observation is based upon the easily verifiable identity
\begin{equation}
\gamma ^0
\Lambda\gamma^0\gamma^\mu
\Lambda
=\gamma^\mu\,.
\end{equation}
Two other useful observations for the evaluation of the above
expressions are the following. The first is
\begin{equation}
\overline{u}^r(p)\gamma^\mu u^s(q)=\frac{1}{2m}\overline{u}^r(p)
(\not{\!p }\gamma^\mu+\gamma^\mu\!\! \not{\! q})u^s(q)
\end{equation}
which uses the fact that $u^r(p)$ is a positive frequency solution
to the Dirac equation; a similar identity can be obtained for
$\overline{v}^r(p)\gamma^\mu u^s(q)$. The second is the observation
that
\begin{equation}
\not{\!p }\gamma^\mu+\gamma^\mu\!\!
\not{\! q }=(p+q)^\mu+i(p-q)_\nu\sigma^{\mu\nu}\,,
\end{equation} where $\sigma^{\mu\nu}=\frac{i}{2}[\gamma^\mu ,\gamma^\nu ]$.
Routine calculations now lead to the following results.
\begin{equation}
\label{terms1}
\begin{array}{ll}
\overline{v}^r(p)\gamma^0 u^s(q)=&\frac{1}{\sqrt{N_pN_q}}\xi^{r\dagger}
\boldsymbol{ \sigma}\cdot(\hat{\pb }+\hat{\qb })
\xi^s\,,\\
\overline{v}^r(p)\gamma^i u^s(q)=&\frac{1}{\sqrt{N_pN_q}}\xi^{r\dagger}
([(\boldsymbol{\sigma}\cdot\hat{\pb })\hat{\qb }+
(\boldsymbol{\sigma}\cdot\hat{\boldsymbol{ q}})\hat{\pb }-i\pb
\times \boldsymbol{ q}]^i+(1-\hat{\pb }\cdot\hat{\qb
})\sigma
^i)\xi^s\,,\\
\overline{u}^r(p)\gamma^0 u^s(q)=&\frac{1}{\sqrt{N_pN_q}}\xi^{r\dagger}
(1+\hat{\pb }\cdot\hat{\qb }+i\boldsymbol{
\sigma}\cdot\hat{\pb }\times\hat{\qb })\xi^s\,, \\
\overline{u}^r(p)\gamma^i u^s(q)=&\frac{1}{\sqrt{N_pN_q}}\xi^{r\dagger}
(\pb +\qb +i\boldsymbol{ \sigma}\times (\hat{\boldsymbol{
p}}-\hat{\qb }))^i\xi^s\,.
\end{array}
\end{equation}
The latter pair of results is sufficient to calculate the
asymptotic interaction Hamiltonian~(\ref{hint_as}) of the massive
theory.

\medskip

In the $m=0$ limit, these results simplify in an attractive way.
Recall from the discussion of Sect.\ 5 that the asymptotic
interaction Hamiltonian only receives a contribution from the
region where the momenta are collinear. As an illustration, we
shall examine the first of the terms in (\ref{terms1}), in the
collinear region where we have $|\kb |=|\qb |+|\kb
-\qb |$ (c.f. the paragraph after Eq.~\ref{hterm2}). In
the massless case we  have $N_p=1/E_p$, so that $\hat{\pb }=\pb
/E_p$. It is also clear that $\hat{\pb }^2=1$ and
$\hat{\pb }\cdot\pb =E_p$. We thus obtain for massless charges for
this momentum configuration
\begin{eqnarray}
\label{apex}
\overline{v}^r(q)\gamma^0
u^s(k-q)&=&\frac{1}{\sqrt{N_{k-q}N_q}}\xi^{r\dagger}
\boldsymbol{\sigma}\cdot(\hat{\qb }+\widehat{(\kb -
\qb)})\xi^s\,, \nonumber\\
&=&
\sqrt{E_qE_{k-q}}\,\xi^{r\dagger}\boldsymbol{\sigma}
\cdot\left(\frac{\qb
}{E_q}+
\frac{\kb -\qb }{E_{k-q}}\right)\xi^s\,.
\end{eqnarray}
Now in the asymptotic interaction Hamiltonian this spinor
combination only occurs if the momenta are such that $k$ and $q$
are parallel light-like vectors, and using this in (\ref{apex}), we
see that
\begin{eqnarray}
\overline{v}^r(q)\gamma^0u^s(k-q)
&=&\sqrt{E_qE_{k-q}}\,\xi^{r\dagger} {\qb }\cdot\boldsymbol{\sigma}
\xi^s\left(\frac{1}{E_q}+\frac{1}{E_q}\right)\\
&=&2\sqrt{E_qE_{k-q}}\,
\xi^{r{\dagger}}\hat{\qb }\cdot\boldsymbol{\sigma}
\xi^s\,.
\end{eqnarray}
The other terms in (\ref{terms1}) can be found in a similar manner.
We confine ourselves to listing them. They yield the simple
results:
\begin{equation}
\label{terms2}
\begin{array}{ll}
\overline{v}^r(q)\gamma^\mu u^s(k-q)=&2\hat{{q}}^\mu
\sqrt{E_qE_{k-q}}\,\xi^{r\dagger}\hat{\qb }\cdot\boldsymbol{\sigma}
\xi^s\\
\overline{u}^r(q)\gamma^\mu u^s(k-q)=&2\hat{p}^\mu
\sqrt{E_qE_{k-q}}\,\delta^{rs}
\end{array}
\end{equation}
These results may be used to extract the asymptotic interaction
Hamiltonian for massless QED.

%\bibliographystyle{h-physrev}
%\bibliography{litbank1}

\begin{thebibliography}{10}

\bibitem{Wigner:1939cj}
E.~P. Wigner,
\newblock Annals Math. {\bf 40}, 149 (1939).

\bibitem{weinberg:1995}
S.~Weinberg,
\newblock {\em The Quantum Theory of Fields} (Cambridge University Press,
  Cambridge, 1995).

\bibitem{Buchholz:1986uj}
D.~Buchholz,
\newblock Phys. Lett. {\bf B174}, 331 (1986).

\bibitem{kulish:1970}
P.~Kulish and L.~Faddeev,
\newblock Theor. Math. Phys. {\bf 4}, 745 (1970).

\bibitem{Lavelle:1997ty}
M.~Lavelle and D.~McMullan,
\newblock Phys. Rept. {\bf 279}, 1 (1997), hep-ph/9509344.

\bibitem{Ciafaloni:1989vs}
M.~Ciafaloni,
\newblock in \textit{Quantum Chromodynamics}, Ed.~A.H.~Mueller (World
  Scientific, Singapore 1989).

\bibitem{DelDuca:1989jt}
V.~D. Duca, L.~Magnea, and G.~Sterman,
\newblock Nucl. Phys. {\bf B324}, 391 (1989).

\bibitem{Henty:1996kv}
UKQCD, D.~S. Henty, O.~Oliveira, C.~Parrinello, and S.~Ryan,
\newblock Phys. Rev. {\bf D54}, 6923 (1996), hep-lat/9607014.

\bibitem{Gogilidze:1997qq}
S.~A. Gogilidze, A.~M. Khvedelidze, D.~M. Mladenov, and H.~P.
Pavel,
\newblock Phys. Rev. {\bf D57}, 7488 (1997), hep-th/9707136.

\bibitem{Maison:1975ex}
D.~Maison and D.~Zwanziger,
\newblock Nucl. Phys. {\bf B91}, 425 (1975).

\bibitem{Frohlich:1979uu}
J.~{Fr\"ohlich}, G.~Morchio, and F.~Strocchi,
\newblock Phys. Lett. {\bf 89B}, 61 (1979).

\bibitem{Zwanziger:1975ka}
D.~Zwanziger,
\newblock Phys. Rev. {\bf D11}, 3504 (1975).

\bibitem{Zwanziger:1975jz}
D.~Zwanziger,
\newblock Phys. Rev. {\bf D11}, 3481 (1975).

\bibitem{Morchio:1983ym}
G.~Morchio and F.~Strocchi,
\newblock Nucl. Phys. {\bf B211}, 471 (1983).

\bibitem{Buchholz:1996}
D.~Buchholz,
\newblock Nucl. Phys. {\bf B469}, 333 (1996), hep-th/9511002.

\bibitem{Dirac:1955ca}
P.~A.~M. Dirac,
\newblock Can. J. Phys. {\bf 33}, 650 (1955).

\bibitem{dEmilio:1984}
E.~d'Emilio and M.~Mintchev,
\newblock Fortschr. Phys. {\bf 32}, 473 (1984).

\bibitem{dEmilio:1984a}
E.~d'Emilio and M.~Mintchev,
\newblock Fortschr. Phys. {\bf 32}, 503 (1984).

\bibitem{Steinmann:1984}
O.~Steinmann,
\newblock Ann. Phys. {\bf 157}, 232 (1984).

\bibitem{Prokhorov:1992}
L.~V. Prokhorov and S.~V. Shabanov,
\newblock Int. J. Mod. Phys. {\bf A7}, 7815 (1992).

\bibitem{Prokhorov:1993}
L.~V. Prokhorov, D.~V. Fursaev, and S.~V. Shabanov,
\newblock Theor. Math. Phys. {\bf 97}, 1355 (1993).

\bibitem{Kawai:1995zx}
T.~Kawai and H.~P. Stapp,
\newblock Phys. Rev. {\bf D52}, 2517 (1995), quant-ph/9502007.

\bibitem{Kawai:1995zv}
T.~Kawai and H.~P. Stapp,
\newblock Phys. Rev. {\bf D52}, 2505 (1995), quant-ph/9511031.

\bibitem{Kawai:1995zu}
T.~Kawai and H.~P. Stapp,
\newblock Phys. Rev. {\bf D52}, 2484 (1995), quant-ph/9503002.

\bibitem{Lusanna:1996ut}
L.~Lusanna and P.~Valtancoli,
\newblock Int. J. Mod. Phys. {\bf A12}, 4769 (1997), hep-th/9606078.

\bibitem{Lusanna:1996us}
L.~Lusanna and P.~Valtancoli,
\newblock Int. J. Mod. Phys. {\bf A12}, 4797 (1997), hep-th/9606079.

\bibitem{Kashiwa:1997}
T.~Kashiwa and N.~Tanimura,
\newblock Phys. Rev. {\bf D56}, 2281 (1997), hep-th/9612250.

\bibitem{Kashiwa:1996}
T.~Kashiwa and N.~Tanimura{, \textit{Physical States and Gauge
Independence of
  the Energy Momentum Tensor in Quantum Electrodynamics}},
\newblock (1996), hep-th/9605207.

\bibitem{Chechelashvili:1997}
G.~Chechelashvili, G.~Jorjadze, and N.~Kiknadze,
\newblock Theor. Math. Phys. {\bf 109}, 1316 (1997), hep-th/9510050.

\bibitem{Haagensen:1997pi}
P.~Haagensen and K.~Johnson{, \textit{On the Wave Functional for
Two Heavy
  Color Sources in Yang-Mills Theory}},
\newblock (1997), hep-th/9702204.

\bibitem{Cahill:1979dq}
K.~Cahill and D.~R. Stump,
\newblock Phys. Rev. {\bf D20}, 540 (1979).

\bibitem{dollard:1964}
J.~Dollard,
\newblock J. Math. Phys. {\bf 5}, 729 (1964).

\bibitem{Bogolyubov:1980nc}
N.~N. Bogolyubov and D.~V. Shirkov,
\newblock {\em Introduction to the Theory of Quantised Fields}, 3rd
Edition, (Wiley-Interscience, New York, 1980).

\bibitem{Lavelle:1996tz}
M.~Lavelle and D.~McMullan,
\newblock Phys. Lett. {\bf B371}, 83 (1996), hep-ph/9509343.

\bibitem{Lavelle:1994xa}
M.~Lavelle and D.~McMullan,
\newblock Phys. Lett. {\bf B329}, 68 (1994), hep-th/9403147.

\bibitem{Bagan:1998kg}
E.~Bagan, M.~Lavelle, and D.~McMullan,
\newblock Phys. Rev. {\bf D57}, 4521 (1998), hep-th/9712080.


\bibitem{Haag:1992hx}
R.~Haag,
\newblock {\em Local Quantum Physics: Fields, Particles, Algebras} (Springer,
  Berlin, 1992).

\bibitem{Georgi:1990um}
H.~Georgi,
\newblock Phys. Lett. {\bf B240}, 447 (1990).

\bibitem{sterman:1993}
G.~Sterman,
\newblock {\em An Introduction to Quantum Field Theory} (Cambridge University
  Press, Cambridge, 1993).

\bibitem{Bagan:1997su}
E.~Bagan, M.~Lavelle, and D.~McMullan,
\newblock Phys. Rev. {\bf D56}, 3732 (1997), hep-th/9602083.

\bibitem{Bagan:1997dh}
E.~Bagan, B.~Fiol, M.~Lavelle, and D.~McMullan,
\newblock Mod. Phys. Lett. {\bf A12}, 1815 (1997), hep-ph/9706515,
\newblock Erratum-ibid, A12 (1997) 2317.

\bibitem{muta:1987}
T.~Muta,
\newblock {\em Foundations of Quantum Chromodynamics} (World Scientific,
  Singapore, 1987).

\bibitem{soloviev:1968}
R.~Jackiw and L.~Soloviev,
\newblock Phys. Rev. {\bf 27}, 1485 (1968).

\bibitem{Dokshitser:1991wu}
Y.~L. Dokshitser, V.~A. Khoze, A.~H. Mueller, and S.~I. Troian,
\newblock {\em Basics of Perturbative QCD} (Ed. Frontieres, Gif-sur-Yvette,
  France, 1991).

\bibitem{Doria:1980ak}
R.~Doria, J.~Frenkel, and J.~C. Taylor,
\newblock Nucl. Phys. {\bf B168}, 93 (1980).

\bibitem{Di'Lieto:1981dt}
C.~Di'Lieto, S.~Gendron, I.~G. Halliday, and C.~T. Sachrajda,
\newblock Nucl. Phys. {\bf B183}, 223 (1981).

\bibitem{Espriu:1996sk}
D.~Espriu and R.~Tarrach,
\newblock Phys. Lett. {\bf B383}, 482 (1996), hep-ph/9604431.

\bibitem{havemann:1985}
F.~Havemann,
\newblock \textit{Collinear Divergences and Asymptotic States{,}} Zeuthen
  Report No. PHE-85-14, 1985 (unpublished),
\newblock scanned at KEK.

\bibitem{Nowakowski:1993iu}
M.~Nowakowski and A.~Pilaftsis,
\newblock Z. Phys. {\bf C60}, 121 (1993), hep-ph/9305321.

\bibitem{jones:1966}
G.S.Jones,
\newblock {\em Generalised Functions}, First ed. (McGraw-Hill, Berkshire,
  1966).

\bibitem{kabat:1992}
R.Jackiw, D.Kabat, and M.Ortiz,
\newblock Phys. Lett. \textbf{B} {\bf 277}, 148 (1992).

\bibitem{robinson:1984}
I.Robinson and D.Rozga,
\newblock J. Math. Phys. {\bf 25}, 499 (1984).

\bibitem{Gribov:1978wm}
V.~N. Gribov,
\newblock Nucl. Phys. {\bf B139}, 1 (1978).

\bibitem{Singer:1978dk}
I.~M. Singer,
\newblock Commun. Math. Phys. {\bf 60}, 7 (1978).

\bibitem{Lavelle:1995rh}
M.~Lavelle and D.~McMullan,
\newblock Phys. Lett. {\bf B347}, 89 (1995), hep-th/9412145.

\bibitem{haller:1997}
L.~Chen, M.~Belloni, and K.~Haller,
\newblock Phys. Rev. {\bf D55}, 2347 (1997), hep-ph/9609507.

\bibitem{Khoze:1997dn}
V.~A. Khoze and W.~Ochs,
\newblock Int. J. Mod. Phys. {\bf A12}, 2949 (1997), hep-ph/9701421.

\bibitem{Parrinello:1991fp}
C.~Parrinello, S.~Petrarca, and A.~Vladikas,
\newblock Phys. Lett. {\bf B268}, 236 (1991).

\bibitem{deForcrand:1995mz}
P.~de~Forcrand and J.~E. Hetrick,
\newblock Nucl. Phys. Proc. Suppl. {\bf 42}, 861 (1995), hep-lat/9412044.

\bibitem{Stuart:1995ba}
R.~G. Stuart,
\newblock in {\em Perspectives for Electroweak Interactions
in e$^+$e$^-$ Collisions}, Ed. B.A. Kniehl, (World Scientific,
Singapore, 1995), hep-ph/9504308.

\bibitem{Papavassiliou:1995fq}
J.~Papavassiliou and A.~Pilaftsis,
\newblock Phys. Rev. Lett. {\bf 75}, 3060 (1995), hep-ph/9506417.

\bibitem{Seiberg:1994rs}
N.~Seiberg and E.~Witten,
\newblock Nucl. Phys. {\bf B426}, 19 (1994), hep-th/9407087.

\bibitem{Lavelle:1993xf}
M.~Lavelle and D.~McMullan,
\newblock Phys. Rev. Lett. {\bf 71}, 3758 (1993), hep-th/9306132.

\bibitem{Bialynicki-Birula:1975yp}
I.~Bialynicki-Birula and Z.~Bialynicka-Birula,
\newblock {\em Quantum Electrodynamics} (Pergamon Press, Oxford, 1975).

\end{thebibliography}

\end{document}